\documentclass[manuscript]{aastex}

\usepackage{amsmath}
\usepackage{threeparttable}
\usepackage{multirow}
\usepackage{lscape}
\usepackage{color}

\slugcomment{Not to appear in Nonlearned J., 45.}
 
\shorttitle{Collisional Disruption with iSALE code}
\shortauthors{Suetsugu et al.}

\begin{document}

%% LaTeX will automatically break titles if they run longer than
%% one line. However, you may use \\ to force a line break if
%% you desire.

\title{Collisional Disruption of Planetesimals in the Gravity Regime with iSALE Code: Comparison with SPH code for Purely Hydrodynamic Bodies}

%% Use \author, \affil, and the \and command to format
%% author and affiliation information.
%% Note that \email has replaced the old \authoremail command
%% from AASTeX v4.0. You can use \email to mark an email address
%% anywhere in the paper, not just in the front matter.
%% As in the title, use \\ to force line breaks.

\author{Ryo Suetsugu\altaffilmark{1}, Hidekazu Tanaka\altaffilmark{2}, Hiroshi Kobayashi\altaffilmark{3}, Hidenori Genda\altaffilmark{4}}
\affil{1. School of Medicine, University of Occupational and Environmental Health,
Iseigaoka, Yahatanishi-ku, KitaKyushu, 807-8555, Japan}
\affil{2. Astronomical Institute, Tohoku University, 
Aramaki, Aoba-ku, Sendai, 980-8578, Japan}
\affil{3. Department of Physics, Graduate School of Science, Nagoya University, 
Furo-cho, Chikusa-ku, Nagoya, 464-8602, Japan}
\affil{4. Earth-Life Science Institute, Tokyo Institute of Technology, 
Ookayama, Meguro-ku, Tokyo, 152-8550, Japan}

\email{ryos@med.uoeh-u.ac.jp}

%% Notice that each of these authors has alternate affiliations, which
%% are identified by the \altaffilmark after each name.  Specify alternate
%% affiliation information with \altaffiltext, with one command per each
%% affiliation.

%% Mark off your abstract in the ``abstract'' environment. In the manuscript
%% style, abstract will output a Received/Accepted line after the
%% title and affiliation information. No date will appear since the author
%% does not have this information. The dates will be filled in by the
%% editorial office after submission.

\begin{abstract}
In most of the previous studies related to collisional disruption of planetesimals in the gravity regime, 
Smoothed Particle Hydrodynamics (SPH) simulations have been used. 
On the other hand, impact simulations using grid-based hydrodynamic code have not been sufficiently performed. 
In the present study, we execute impact simulations in the gravity regime using the shock-physics code iSALE, which is a grid-based Eulerian hydrocode. 
We examine the dependence of the critical specific impact energy $Q_{\rm RD}^{*}$ on impact conditions for a wide range of specific impact energy ($Q_{\rm R}$) from disruptive collisions to erosive collisions, and compare our results with previous studies. 
We find collision outcomes of the iSALE simulation agree well with those of the SPH simulation.
Detailed analysis mainly gives three results.
(1) The value of $Q_{\rm RD}^{*}$ depends on numerical resolution, and  is close to convergence with increasing numerical resolution.
The difference in converged value of $Q_{\rm RD}^{*}$ between the iSALE code and the SPH code is within 30\%.
(2) Ejected mass normalized by total mass ($M_{\rm ej}/M_{\rm tot}$) generally depends on various impact conditions. 
However, when $Q_{\rm R}$ is normalized by $Q_{\rm RD}^{*}$ that is calculated for each impact simulation, 
$M_{\rm ej}/M_{\rm tot}$  can be scaled by $Q_{\rm R}/Q_{\rm RD}^{*}$, and is independent of numerical resolution, impact velocity and target size. 
(3) This similarity law for $Q_{\rm R}/Q_{\rm RD}^{*}$ is confirmed for a wide range of specific impact energy. 
We also derive a semi-analytic formula for  $Q_{\rm RD}^{*}$ based on the  similarity law and the crater scaling law.  
We find that the semi-analytic formula for the case with a non-porous object is consistent with numerical results. 
\end{abstract}

%% Keywords should appear after the \end{abstract} command. The uncommented
%% example has been keyed in ApJ style. See the instructions to authors
%% for the journal to which you are submitting your paper to determine
%% what keyword punctuation is appropriate.


%% From the front matter, we move on to the body of the paper.
%% In the first two sections, notice the use of the natbib \citep
%% and \citet commands to identify citations.  The citations are
%% tied to the reference list via symbolic KEYs. The KEY corresponds
%% to the KEY in the \bibitem in the reference list below. We have
%% chosen the first three characters of the first author's name plus
%% the last two numeral of the year of publication as our KEY for
%% each reference.

%% Authors who wish to have the most important objects in their paper
%% linked in the electronic edition to a data center may do so by tagging
%% their objects with \objectname{} or \object{}.  Each macro takes the
%% object name as its required argument. The optional, square-bracket 
%% argument should be used in cases where the data center identification
%% differs from what is to be printed in the paper.  The text appearing 
%% in curly braces is what will appear in print in the published paper. 
%% If the object name is recognized by the data centers, it will be linked
%% in the electronic edition to the object data available at the data centers  
%%
%% Note that for sources with brackets in their names, e.g. [WEG2004] 14h-090,
%% the brackets must be escaped with backslashes when used in the first
%% square-bracket argument, for instance, \object[\[WEG2004\] 14h-090]{90}).
%%  Otherwise, LaTeX will issue an error. 

\section{INTRODUCTION}
Collisions are one of the most important processes in planet formation because planetary bodies in the Solar System are thought to have experienced a lot of collisions 
during the accretion process \citep[e.g.,][]{L93}. 
Thus, collisional processes have been examined extensively. 
Roughly speaking, collisional outcomes can be classified into disruptive collisions and erosive collisions by the specific impact energy $Q_{\rm R}$, given by 
\begin{eqnarray} 
Q_{\rm R}=\left(\frac{1}{2}M_{\rm tar}V_{\rm tar}^{2}+\frac{1}{2}M_{\rm imp}V_{\rm imp}^{2}\right)/M_{\rm tot}=\left(\frac{1}{2}M_{\rm R} v_{\rm imp}^{2}\right)/M_{\rm tot},
\end{eqnarray}
where $M_{\rm tar}$ and $M_{\rm imp}$ are the mass of the target and the impactor ($M_{\rm tar}>M_{\rm imp}$, $M_{\rm tot}=M_{\rm tar}+M_{\rm imp}$), respectively, 
and $V_{\rm tar}$ and $V_{\rm imp}$ are the velocities of the target and the impactor in the frame of the center of mass when the two objects contact each other, respectively,  
$M_{\rm R}$ is the reduced mass, given by $M_{\rm imp}M_{\rm tar}/M_{\rm tot}$, and $v_{\rm imp}$ is the impact velocity ($v_{\rm imp}=V_{\rm imp}-V_{\rm tar}$ for negative $V_{\rm tar}$). 
In particular, the specific impact energy required to disperse the largest body such that it has exactly half its total mass after the collision is called the critical specific impact energy $Q_{\rm RD}^{*}$.
In the case of $Q_{\rm R}>Q_{\rm RD}^{*}$, collisions between planetesimals are regarded as disruptive collisions, 
while they are non-disruptive collisions for $Q_{\rm R}\ll Q_{\rm RD}^{*}$, whose mass ejection is small (hereafter called erosive collisions). 
 
The values of $Q_{\rm RD}^{*}$ have been investigated by laboratory experiments and numerical simulations \citep[e.g.,][]{BA99, N09}. 
When the target is small enough to neglect the effect of the target's gravity, the critical specific impact energy is mainly estimated by laboratory experiments \citep{HH99, N09}. 
As target size increases, collision outcomes gradually become dominated by the gravity of the target.
However, direct experimental measurements of a large scale collision are difficult to carry out in the laboratory. 
Thus, the values of $Q_{\rm RD}^{*}$ for large targets ($\gtrsim 1$km) are estimated via shock-physics code calculations, which compute the propagation of the shock wave caused by a high velocity collision ($\gtrsim$ km/s): Lagrangian hydrocode such as Smoothed Particle Hydrodynamics (SPH) methods \citep{LA96, MR97, BA99, J10, G15, J15, M16, G17}, or a hybrid code of Eulerian hydrocode and N-body \citep{LS09}. 
These numerical simulations showed the dependence of the value of $Q_{\rm RD}^{*}$ on various impact conditions such as target size, impact velocity, material properties, and impact angle. 
For example, the value of $Q_{\rm RD}^{*}$ in the gravity regime increases nearly monotonically with the size of the target 
because collisional fragments are more easily bound by the gravitational force of the target.
The critical specific impact energy also depends on the material property (e.g. material strength, porosity, and friction) of the impactor and the target \citep{LS09, J10, J15}. 
Notably, the friction significantly dissipates impact energy \citep{KG18}, which tends to hinder the disruption of the target. 
The value of $Q_{\rm RD}^{*}$ then reaches about 10 times the value of $Q_{\rm RD}^{*}$ without the friction \citep{J15}. 
Moreover, recent impact simulations show that $Q_{\rm RD}^{*}$ depends not only on impact conditions, but also on numerical resolution \citep{G15,G17}. 
\citet{G15} performed SPH simulation at various numerical resolutions, 
and showed that $Q_{\rm RD}^{*}$ at high numerical resolution is rather low compared to the case of low resolution. 

In addition to the critical specific impact energy, the understanding of erosive collisions is also important in relation to the formation of planetary bodies. 
In most of the previous studies, the contribution of erosive collision to growth of the planets has been underestimated because the amount of mass ejected by erosive collision is much smaller than the total mass. 
However, some previous studies showed that erosive collision also plays an important role in planetary accretion \citep{KT10, K10, K11}.  \citet{KT10} assumed  a simple fragmentation model describing both disruptive collisions and erosive collisions, and investigated mass depletion time in a collision cascade based on analytic consideration and numerical simulation. 
They showed that erosive collisions occur much more frequently than disruptive collisions and the mass depletion time is mainly determined by erosive collisions. 
Recently, the validity of the simple fragmentation model was examined by \citet{G17}, who performed impact simulations for a wide range of specific impact energy using the SPH method with self-gravity and without material strength (i.e. a purely hydrodynamic case), 
and showed that the fragmentation model is consistent with collisional outcomes of simulations within a factor of two. 
They also showed that the ejected mass normalized by the total mass can be scaled by $Q_{\rm R}/Q_{\rm RD}^{*}$ for their parameter range.  

However, almost all high velocity collisions have been examined by the SPH method. 
Another common hydrodynamic simulation, whose computational domain is discretized by grids, has also been carried out \citep[e.g.,][]{LS09}.
However, the grid-based code is only used for the shock deformation immediately after collision, 
and a large part of the disruption is calculated by N-body simulation. 
Thus, impact simulation using the grid-based code has not been sufficiently performed, though it is important to examine the problem with a different numerical approach.

In this study, we perform impact simulations in the gravity regime by using shock-physics code iSALE \citep{A80, C04, W06, C16}, 
which is a grid-based Eulerian hydrocode, and has been widely distributed to academic users in the impact community.
This code has been used to understand various impact phenomena: 
crater formation \citep{C08,C12}, impact jetting \citep{JM15, W17, K18}, pairwise collisions of planetesimals with/without self-gravity \citep{D10, D12} and comparison with experimental data \citep{N16, KA18}.  
We examine the dependence of $Q_{\rm RD}^{*}$ on numerical resolution and impact conditions for a wide range of specific impact energy from disruptive collisions to erosive collisions, and compare our results with previous studies. 
Furthermore, using numerical results obtained by the iSALE code and the crater scaling law, we derive a semi-analytic formula for $Q_{\rm RD}^{*}$. 

In Section~\ref{sec:num}, we present methods for impact simulations and analysis.
We show  our numerical outcomes of simulations in the case of disruptive collisions in Section~\ref{sec:result1}.
In Section~\ref{sec:result2}, we establish a similarity law for $Q_{\rm R}/Q_{\rm RD}^{*}$ for a wide range of impact energy, 
and derive a semi-analytic formula for $Q_{\rm RD}^{*}$.
We discuss effects of oblique collisions and material properties in our results in Section~\ref{sec:dis}.
Section~\ref{sec:sum} summarizes our results.

%%%%%%%%%%%%%%%%%%%%%%%%%%%%%%%%%%%%%%%%%%%%%%%%%%%%%%%%%%%%%%%%%%%%%%
%%%%%%%%%%%%%%%%%%%%%%%%%%%%%%%%%%%%%%%%%%%%%%%%%%%%%%%%%%%%%%%%%%%%%%
\section{NUMERICAL METHODS}
\label{sec:num}
In this study, we examine  collisions between planetesimals using shock-physics code iSALE-2D, 
the version of which is iSALE-Chicxulub.
The iSALE-2D is an extension of the SALE hydrocode \citep{A80}. 
To simulate hypervelocity impact processes in solid materials, SALE was modified to include an elasto-plastic constitutive model, fragmentation models, and multiple materials \citep{M92, I97}. 
More recent improvements include a modified strength model \citep{C04}, and a porosity compaction model \citep{W06, C11}.

The iSALE-2D supports two types of equation of state: ANEOS \citep{TL72, M07} and Tillotson equation of state \citep{T62}. 
These equations of state have been widely applied in previous studies
including planet- and planetesimal-size collisional simulations \citep[e.g.][]{CA01, C4, F10, CS12, SG12, H16, W17}.
In our simulation, we use the Tillotson equation of state for basalt because almost all previous studies related to collisional disruption have used the Tillotson equation of state, 
which allows us to directly compare our results with theirs.
The Tillotson equation of state contains ten material parameters, and the pressure is expressed as a function of the density and the specific internal energy;
all of which are convenient when used in works regarding fluid dynamics. 
Although the Tillotson parameters for basalt of the iSALE-2D are set to experimental values,  
we used the parameter sets of basalt referenced in previous works \citep{BA99, G15, G17}.

\begin{figure}[ht]
 \begin{center}
  \includegraphics[width=85mm]{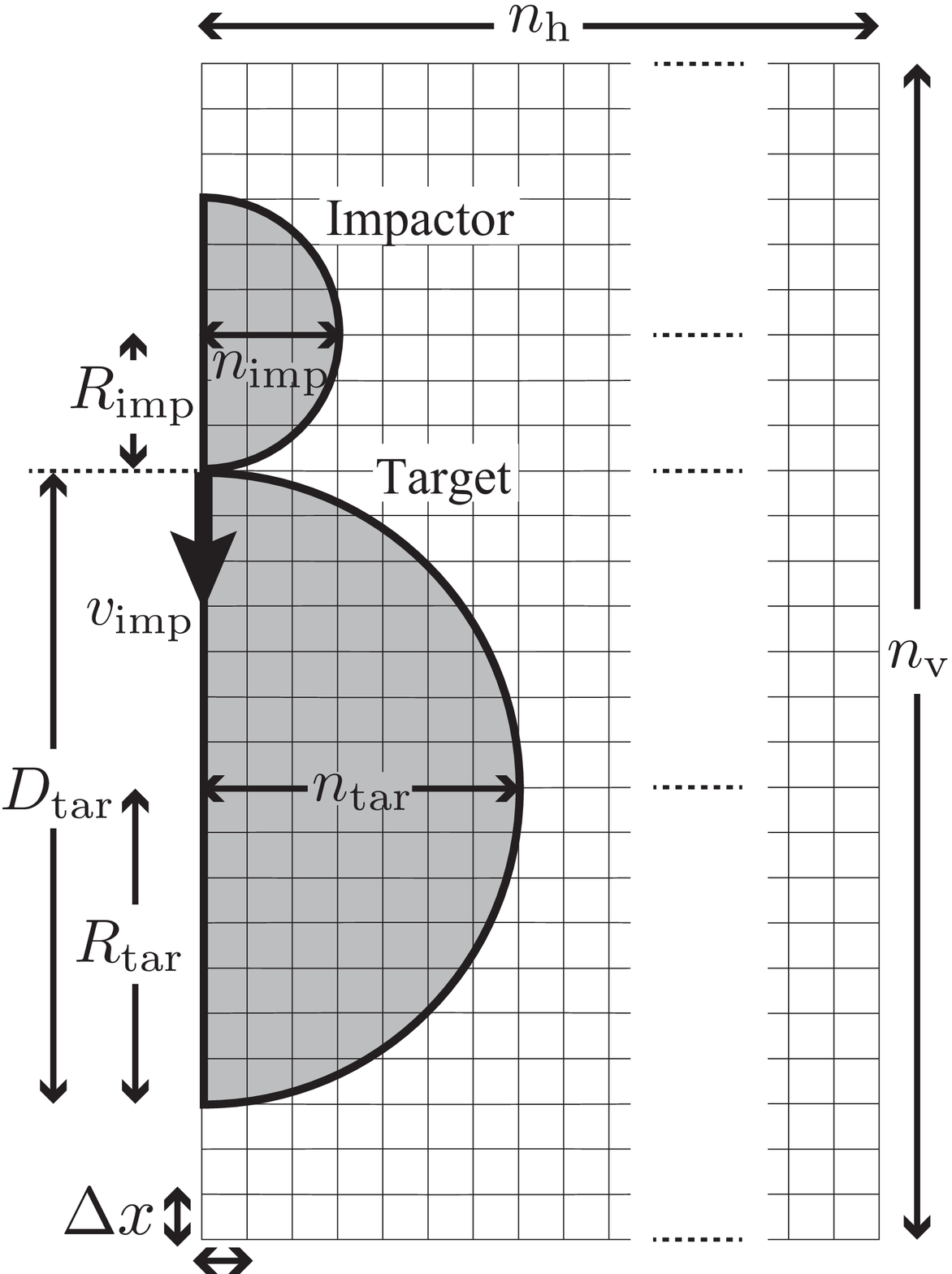}
 \end{center}
 \caption{Schematic illustration of our numerical setting. 
In a two-dimensional cylindrical coordinate system, 
an impactor with velocity $v_{\rm imp}$ has a head-on collision with a target.
The radii of the target and impactor are $R_{\rm tar}$ and $R_{\rm imp}$, respectively.}
 \label{fig:illust}
\end{figure}

We  employ the two-dimensional cylindrical coordinate system and perform head-on impact simulations between two planetesimals (Figure~\ref{fig:illust}).
We assumed that planetesimals are not differentiated. 
Planetesimals are also assumed to be composed of basalt.
For nominal cases, the radius of the target $R_{\rm tar}$ and the impact velocity of the impactor $v_{\rm imp}$ are fixed at 100 km and 3 km/s, respectively.
We also examine the dependence of collisional outcome on target size and impact velocity in Section~\ref{subsec:vimp_rtar}.
To carry out impact simulations with various impact energy $Q_{\rm R}$,  we changed the radius of the impactor $R_{\rm imp}$.
For example, $R_{\rm imp}=14$ - 21 km (i.e., $Q_{\rm R}\simeq 12$ - 41 kJ/kg).
In this study, we consider four cases with the number of cells per target radius ($n_{\rm tar} = 100, 200, 400$, and $800$).
Then, the total number of numerical cells in the computational domain ($n_{\rm v}\times n_{\rm h}$, see Figure~\ref{fig:illust}) is changed depending on $n_{\rm tar}$. 
For example,  $(n_{\rm v}\times n_{\rm h})=(450\times 450), (900\times900), (1800\times1800)$, and $(3600\times3600)$  at $n_{\rm tar}=100$, 200, 400, and 800, respectively.
In the case of $R_{\rm tar}=100$ km, the values of the spatial cell size for each numerical resolution $\Delta x(=R_{\rm tar}/n_{\rm tar})$ are $\Delta x=1000$, 500, 250, and 125 m, 
and the size of the computational domain is fixed at $(n_{\rm v}\Delta x, n_{\rm h}\Delta x)=$(450 km, 450 km). 

The aim of this study is to make a direct comparison of collisional outcomes between different numerical codes (SPH code and iSALE code).
Therefore, although the iSALE-2D can deal with the effects of material strength, damage, and porosity of the target and the impactor, 
these effects are not taken into account in the present work; that is, the fluid motion is purely hydrodynamic.
The self-gravity is calculated by the algorithm in the iSALE-2D based on a Barnes-Hut type algorithm, which can reduce the computational cost of updating the gravity field.
In most of our calculations, the opening angle $\theta$, which is the ratio of mass length-scale to separation distance, is set to 1.0.
Although the value of the opening angle is rather large, we confirmed that the difference in ejected mass between cases for $\theta=1.0$ and $\theta=0.5$ (or 0.1) is within $\sim$5 \% for impacts of our interest. 
For example, we compared the results for $R_{\rm tar}=300$ km because the  gravity of such a large planetesimal is relatively strong (Section~\ref{subsec:vimp_rtar}).
Then, the calculation time of a single impact simulation mainly depends on the numerical resolution.
For example, in our calculation with $n_{\rm tar}=800$,
the calculation time is a few weeks for $R_{\rm tar}=100$ km, while it is two months for $R_{\rm tar}=30$ km, using a computer with an Intel$^{\rm R}$ Core$^{\rm TM}$ i7-4770K  Processor (3.50 GHz).
Other input parameters for iSALE simulation are summarized in Table~\ref{table:input}.

In order to obtain the value of $Q_{\rm RD}^{*}$, we need to estimate the mass of ejecta caused by a single disruptive collision, 
$M_{\rm ej}$, defined as  
\begin{eqnarray}
M_{\rm ej}={M_{\rm tot}}-M_{\rm lrg},
\label{eq:def_mej}
\end{eqnarray}
where $M_{\rm lrg}$ is the mass of the largest body resulting from the impact.
The mass of the largest body $M_{\rm lrg}$ is determined by the following procedure.
First, we find the groups of cells in contact with cells of non-zero densities and 
compare the total masses of the cells in the respective groups. 
The largest total mass is regarded as the preliminary mass of the largest body $M_{\rm lrg}$.
This procedure roughly corresponds to a friends-of-friends algorithm to identify collisional fragments used in previous studies with a SPH method \citep[e.g.,][]{BA99, G15, G17}.
Next, if the constituent cells in the tentative largest body are gravitationally bound, the largest body is determined. 
Otherwise, we find the cells where the specific kinetic energies are larger than the specific gravitational potential energy of the tentative largest body 
and remove them from the largest body.
This procedure is iteratively performed until the number of the cells in the largest mass converges. 
We regard the converged $M_{\rm lrg}$ as the mass of the largest body.

%%%%%%%%%%%%%%%%%%%%%%%%%%%%%%%%%%%%%%%%%%%%%%%%%%%%%%%%%%%%%
%%%%%%%%%%%%%%%%%%%%%%%%%%%%%%%%%%%%%%%%%%%%%%%%%%%%%%%%%%%%%

\section{DISRUPTIVE COLLISIONS AND THE CRITICAL SPECIFIC IMPACT ENERGY $Q_{\rm RD}^{*}$}
\label{sec:result1}
\subsection{$Q_{\rm RD}^{*}$ at a nominal case}
\label{subsec:reso}
\begin{figure}[ht]
 \begin{center}
  \includegraphics[width=50mm]{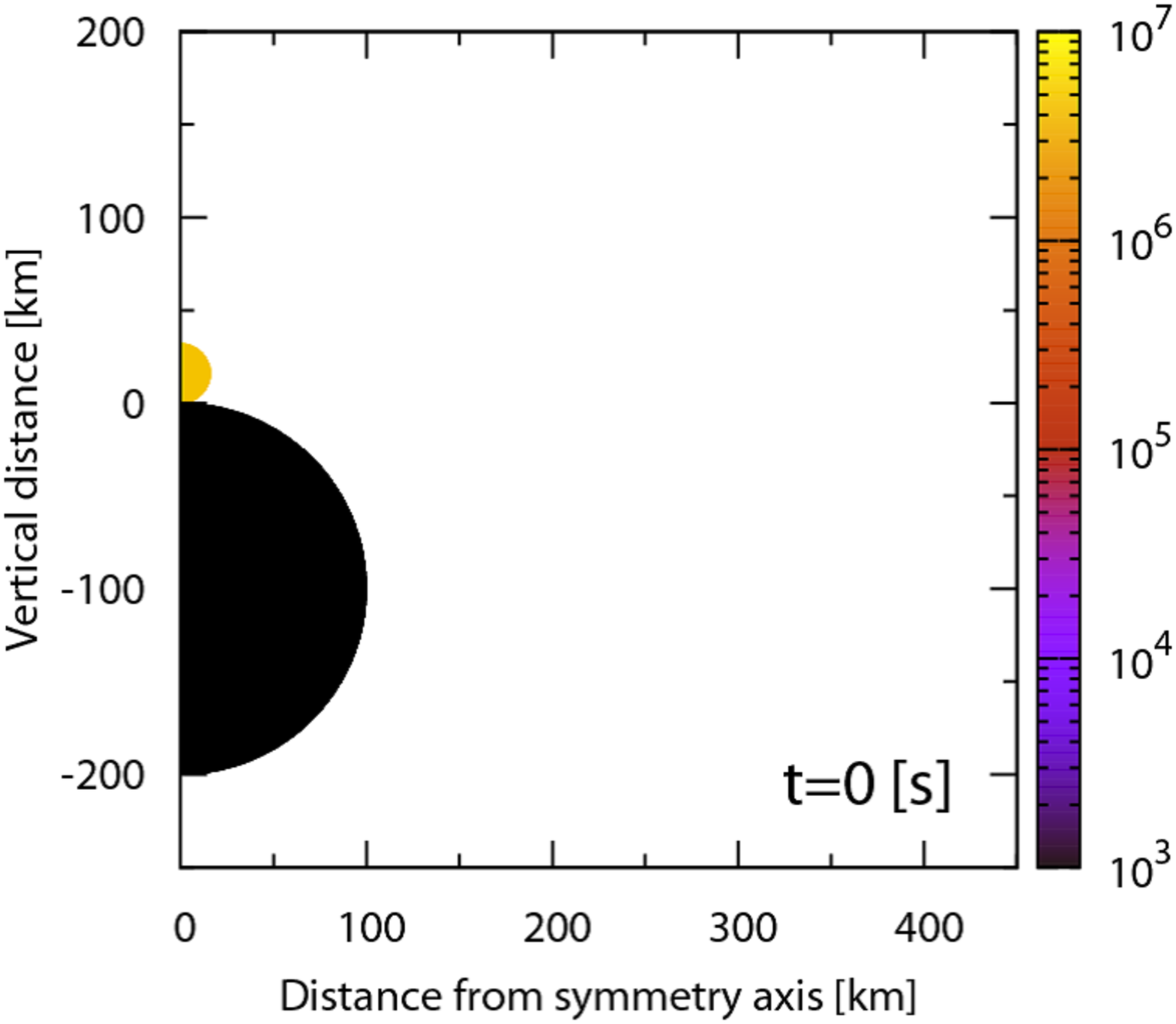}
  \includegraphics[width=50mm]{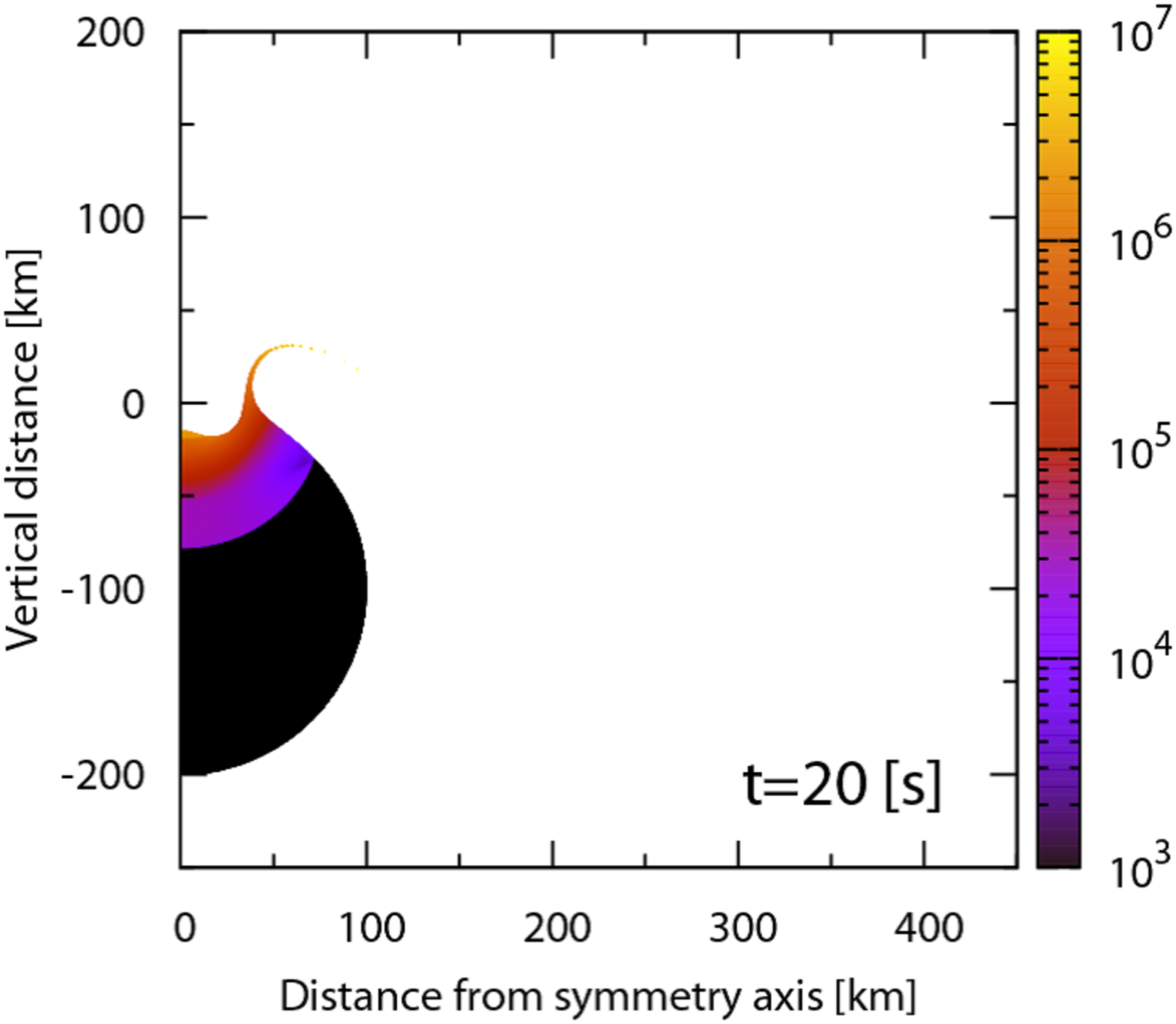}
  \includegraphics[width=50mm]{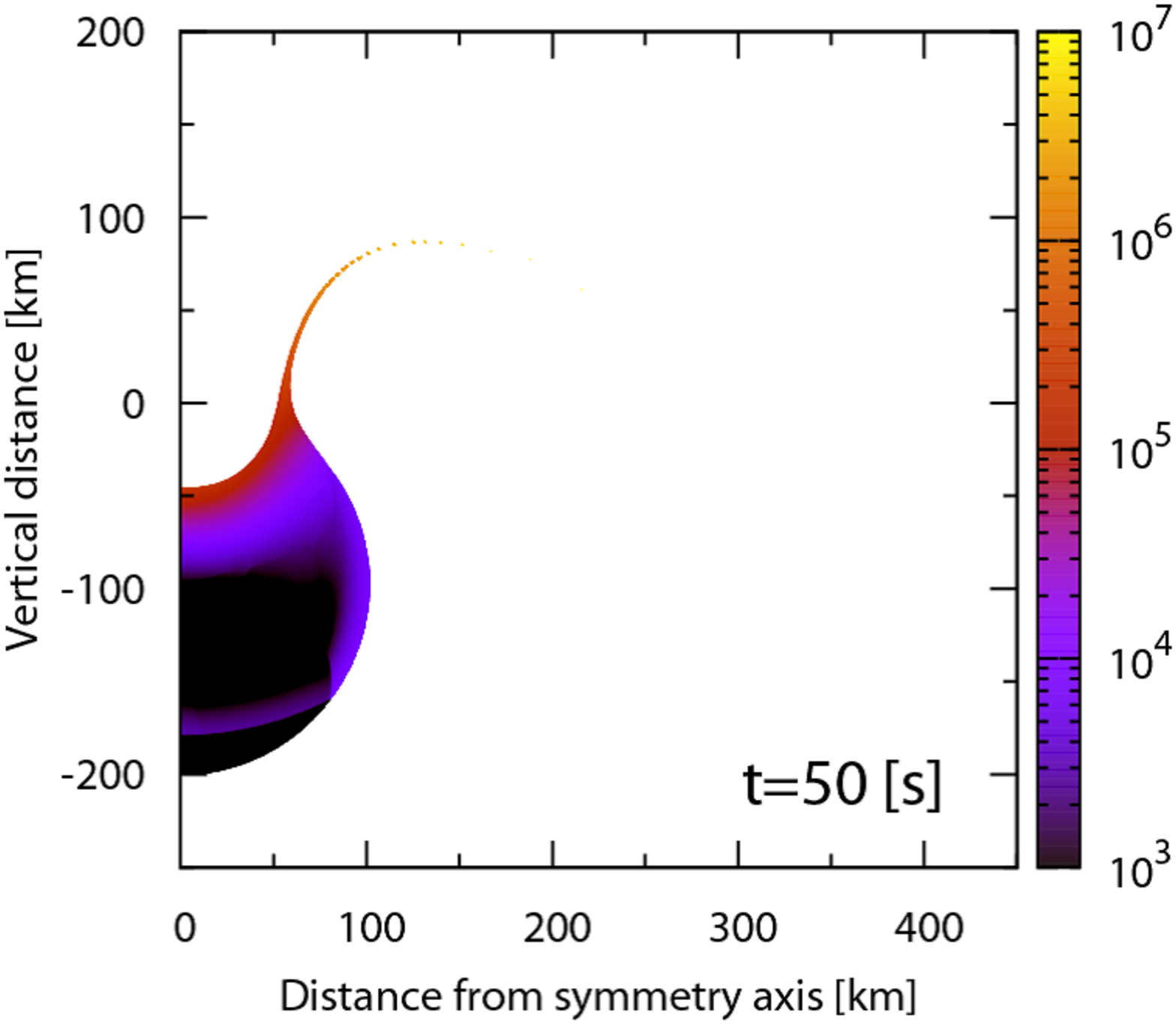}
  \includegraphics[width=50mm]{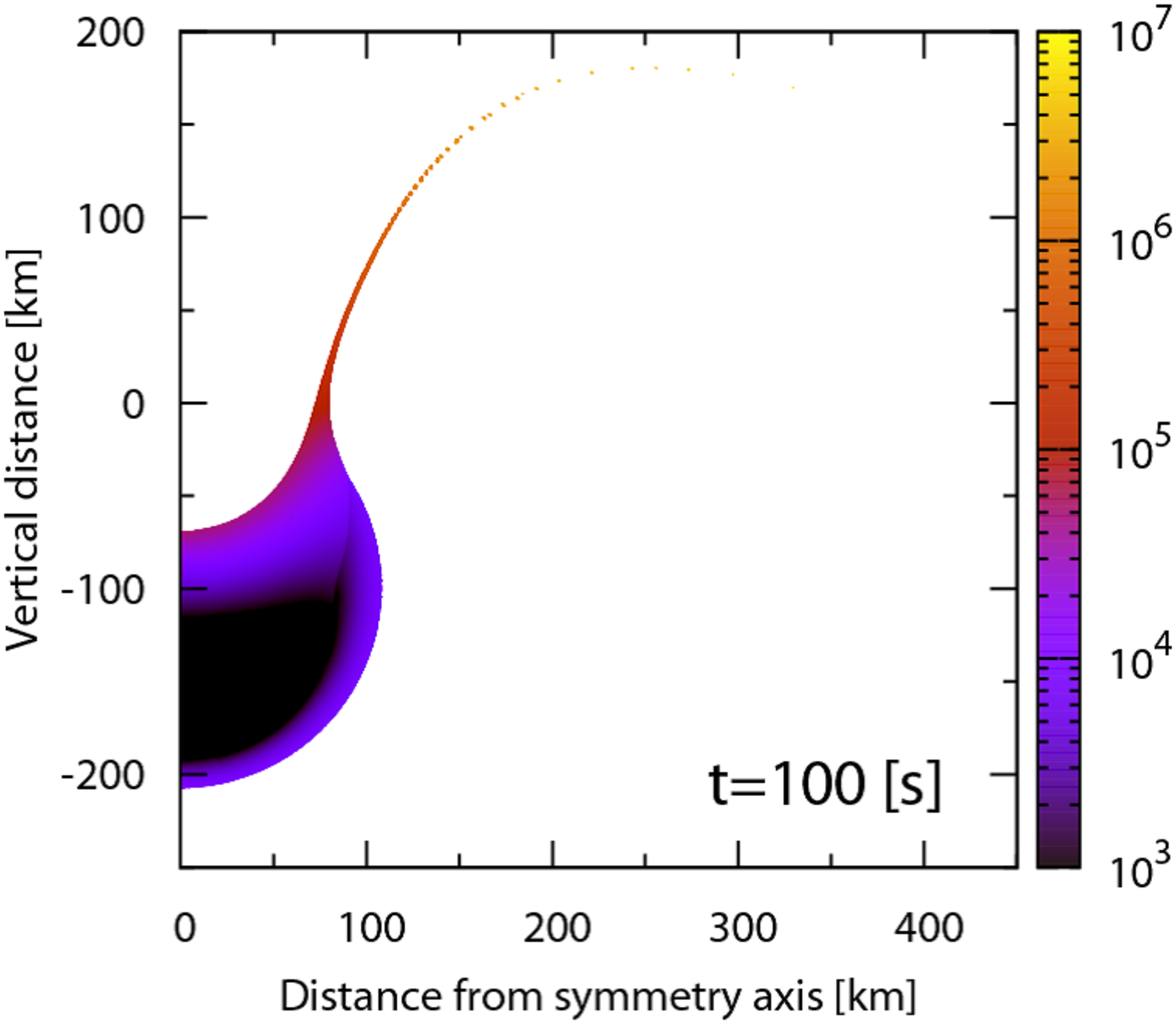}
  \includegraphics[width=50mm]{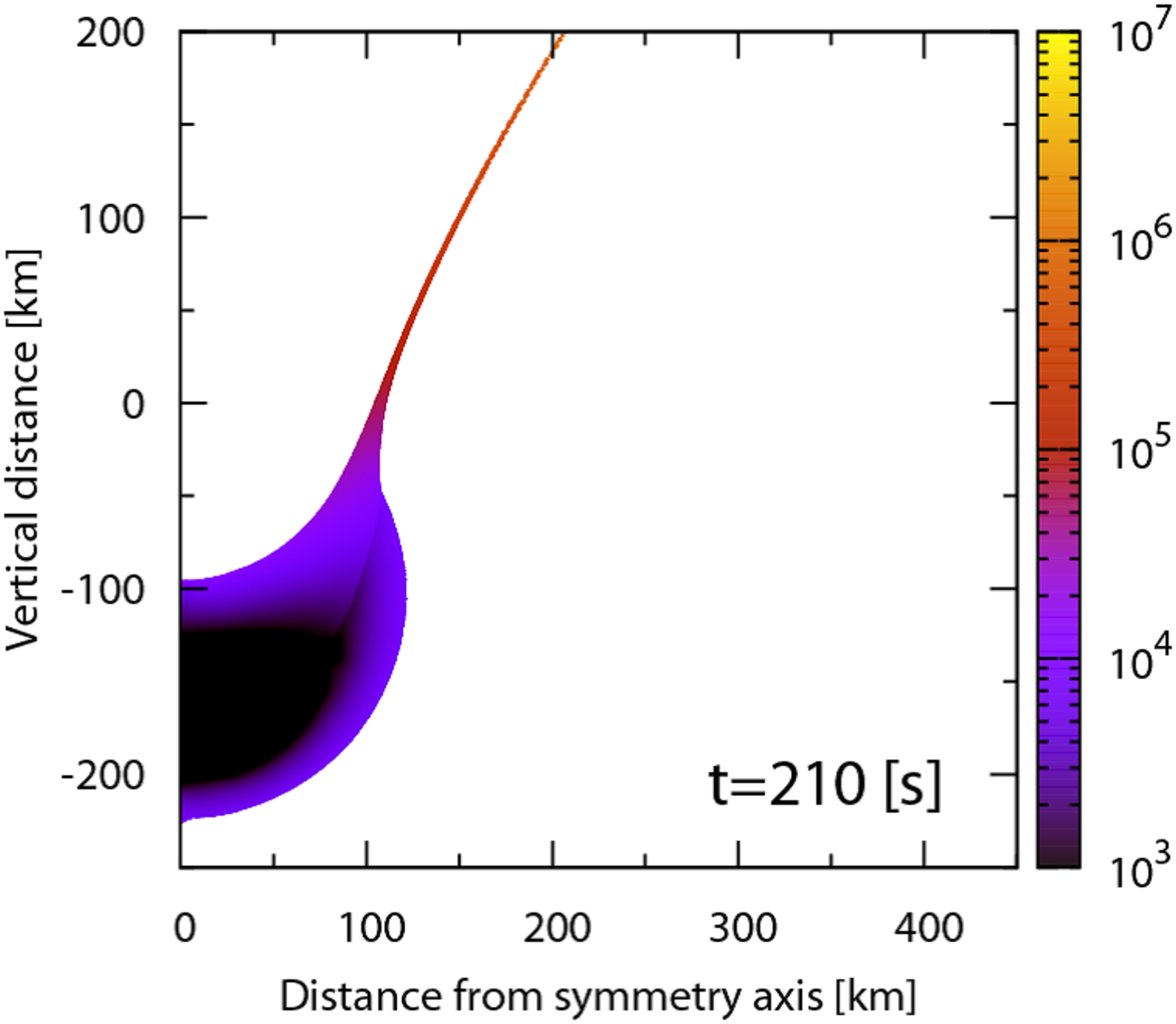}
  \includegraphics[width=50mm]{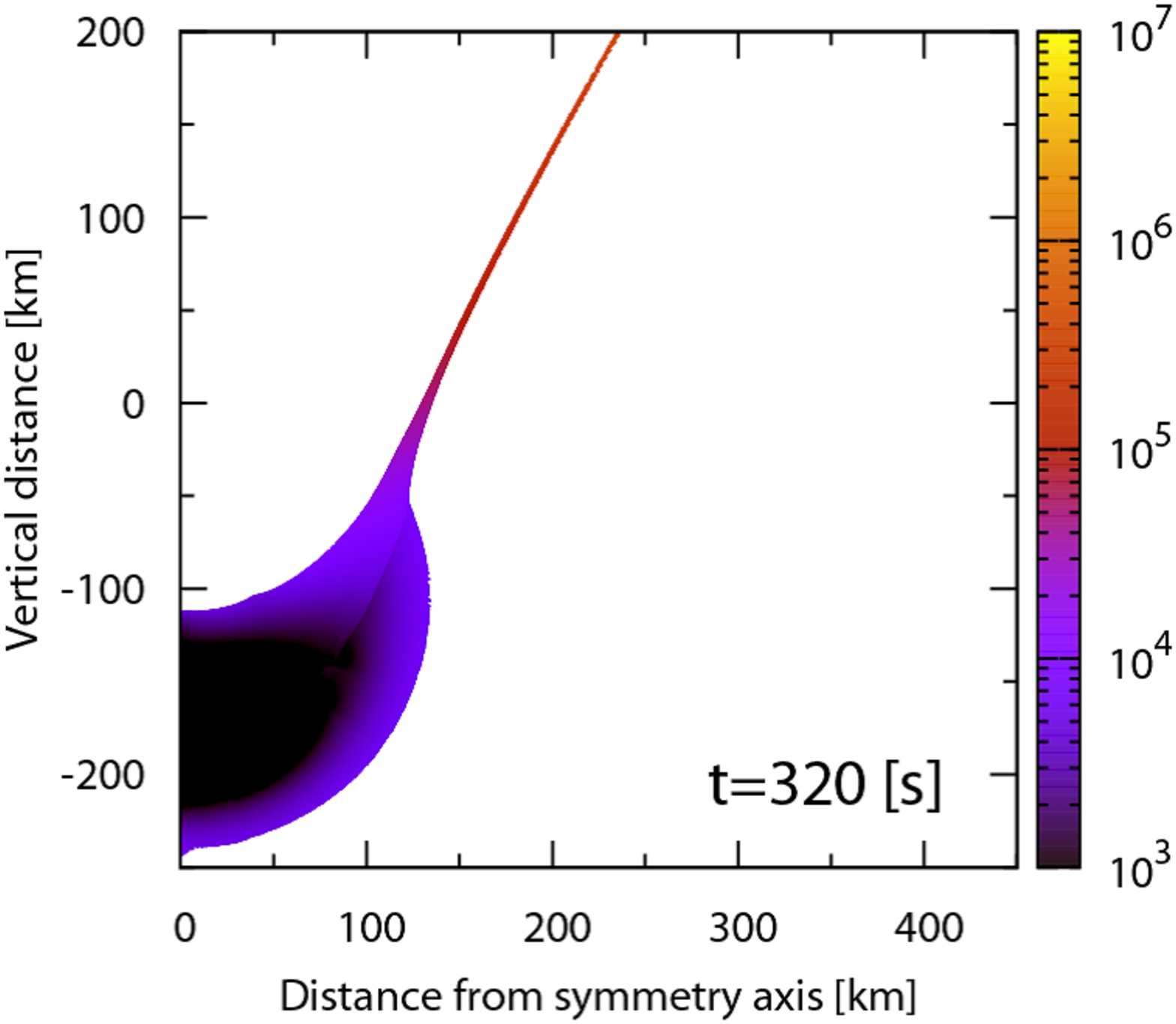}
  \end{center}
 \caption{
Time series of a simulation of a head-on impact between a target with $R_{\rm tar}=100$ km  and an impactor with $R_{\rm imp}=16$ km.
The impact velocity and the number of cells per target radius are set to be 3 km/s, and  800, respectively. 
The color contour represents the specific kinetic energy [J/kg].
}
 \label{fig:snapshot}
\end{figure}

We examine the critical specific impact energy $Q_{\rm RD}^{*}$ for a nominal case with the target radius $R_{\rm tar}=100$ km and the impact velocity $v_{\rm imp}=3$ km/s.
Figure~\ref{fig:snapshot} shows a time series of a simulation of an impactor's head-on collision ($R_{\rm imp}=16$ km) with the target ($R_{\rm tar}=100$ km) at an  impact velocity of $v_{\rm imp}=3$ km/s. 
We adopt the number of cells per target radius $n_{\rm tar}=800$.
The color contour represents the specific kinetic energy.
When $t=0$ s, the impactor starts colliding with the target.
The shock wave generated by the impact propagates through the target and arrives at the antipode of the impact point at $t\simeq50$ s.
Most of the ejecta formed by the collision have relatively large amounts of kinetic energy, and continue escaping from the gravity of the target even when $t\gtrsim300$ s.

\begin{figure}[ht]
 \begin{center}
  \includegraphics[width=80mm]{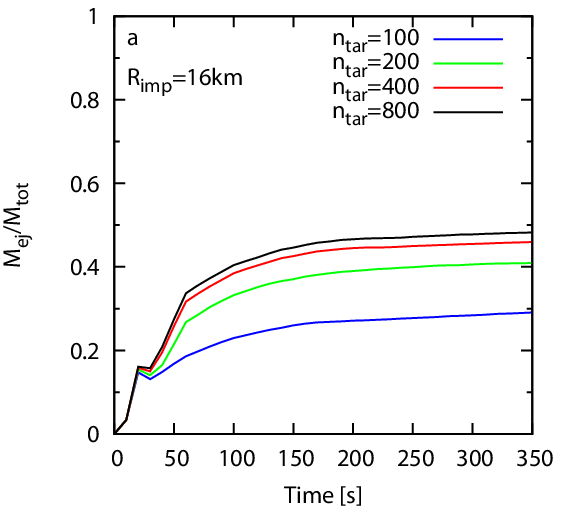}
  \includegraphics[width=80mm]{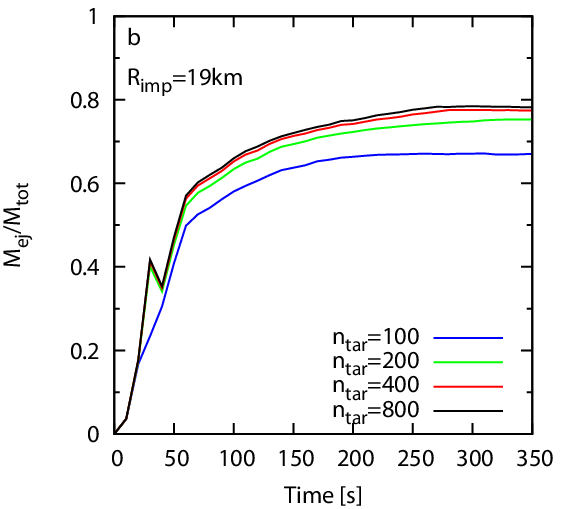}
 \end{center}
 \caption{
The time evolution of ejected mass normalized by the total mass ($M_{\rm ej}/M_{\rm tot}$) for $R_{\rm imp}=16$ km (a) and 19 km (b).
Blue, green, red and black represent the cases with $n_{\rm tar}=100$, 200, 400, and 800, respectively.
Black curve in Panel (a) corresponds to the collision shown in Figure~\ref{fig:snapshot}.
}
 \label{fig:tmlr}
\end{figure}

In Figure~\ref{fig:tmlr}(a), the black curve represents the time evolution of the ejected mass ($M_{\rm ej}$) normalized by the total mass of colliding bodies ($M_{\rm tot}$) in the impact simulation shown in Figure~\ref{fig:snapshot}.
Immediately after the collision, the value of $M_{\rm ej}/M_{\rm tot}$ increases rapidly.
However, it converges within a short time ($t\lesssim170$ s) and eventually becomes nearly constant ($M_{\rm ej}/M_{\rm tot}\simeq0.49$) at $t=350$ s. 
Figure~\ref{fig:tmlr}(a) also shows the resolution dependence of the time evolution of $M_{\rm ej}/M_{\rm tot}$.
The general feature of the ejected mass  for each numerical resolution is similar to that with $n_{\rm tar}=800$. 
However, the converged values of $M_{\rm ej}/M_{\rm tot}$ increase as numerical resolution increases, which was also observed in the results obtained by using the SPH code \citep{G15}.
Figure~\ref{fig:tmlr}(b) shows the case with more destructive collision ($R_{\rm imp}=19$ km). 
Although the differences in the values of $M_{\rm ej}/M_{\rm tot}$ between numerical resolutions become smaller, 
the basic feature is similar to the cases with $R_{\rm imp}=16$ km.

\begin{figure}[ht]
 \begin{center}
 \includegraphics[width=80mm]{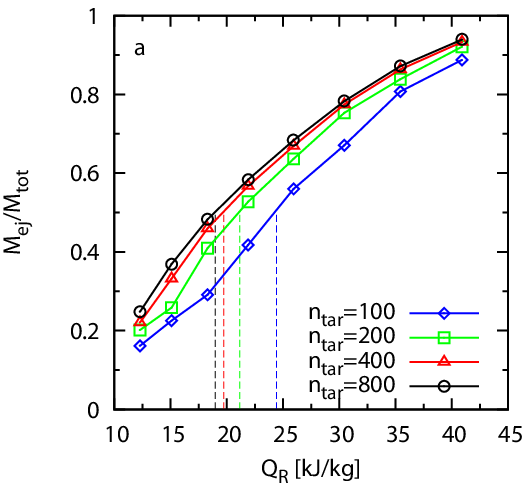}
 \includegraphics[width=80mm]{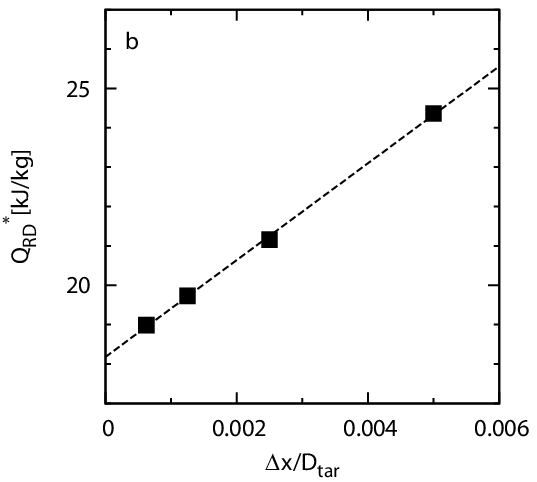}
  \end{center}
 \caption{
Determination of $Q_{\rm RD}^{*}$ value and its dependence on the numerical resolution.
Panel (a) shows normalized mass of ejecta due to disruptive collisions as a function of the impact energy.
Diamonds, squares, triangles, and circles represent the cases with $n_{\rm tar}=100$, 200, 400, and 800, respectively.
The data listed in Table~\ref{table:output} are used.
The vertical dashed lines represent the critical specific impact energies $Q_{\rm RD}^{*}$ for  each resolution.
Panel (b) shows the critical specific impact energy $Q_{\rm RD}^{*}$ as a function of the spatial cell size divided by the target's diameter ($\Delta x/D_{\rm tar}$). 
The dashed line is the least-squares fit to the results (Equation~(\ref{eq:fit})).
}
 \label{fig:qd_m100}
\end{figure}

Figure~\ref{fig:qd_m100}(a) shows $M_{\rm ej}/M_{\rm tot}$ for various numerical resolutions as a function of $Q_{\rm R}$. 
The values of $M_{\rm ej}/M_{\rm tot}$ are listed in Table~\ref{table:output}.
In all the cases, we find that the ejected mass increases as impact energy increases. 
We also find that $M_{\rm ej}/M_{\rm tot}$  for each impact energy increases with $n_{\rm tar}$, 
and the differences in $M_{\rm ej}/ M_{\rm tot}$ between numerical resolutions become smaller in the case of higher resolution. 
The vertical dashed lines in Figure~\ref{fig:qd_m100}(a) represent the critical specific impact energy $Q_{\rm RD}^{*}$ for  each numerical resolution, 
which can be calculated by the linear interpolation of the two data sets of $Q_{\rm R}$ across $M_{\rm ej}/M_{\rm tot}=0.5$  \citep[see][]{G15,G17}.
We find $Q_{\rm RD}^{*}=24.37, 21.16, 19.73$ and 18.99 kJ/kg for $n_{\rm tar}=100$, 200, 400, and 800, respectively.
Figure~\ref{fig:qd_m100}(b) shows $Q_{\rm RD}^{*}$ as a function of  the spatial cell size $\Delta x$ divided by the diameter of the target $D_{\rm tar}$.
Here, $\Delta x/D_{\rm tar}$ corresponds to  $(2n_{\rm tar})^{-1}$.
The values of $Q_{\rm RD}^{*}$ decrease monotonically with decreasing $\Delta x/D_{\rm tar}$ and would be close to convergence.
Even for the highest resolution simulation ($n_{\rm tar}=800$), the value of $Q_{\rm RD}^{*}$ does not fully converge.
However, \citet {G15, G17} showed the dependence of  $Q_{\rm RD}^{*}$ can be approximated well by a linear function of the inverse of the numerical resolution. 
Therefore, we also  fit  our result with the following relation,
\begin{eqnarray}
Q_{\rm RD}^{*} = Q_{\rm RD,0}^{*}+b(\Delta x/D_{\rm tar}),
\label{eq:fit}
\end{eqnarray}
where $Q_{\rm RD,0}^{*}$ and $b$ are fitting parameters.
As a result, we find that the fitting parameters are determined to be $Q_{\rm RD,0}^{*}=18.25$ kJ/kg and $b=1241$ kJ/kg in the case of $R_{\rm tar} = 100$ km and $v_{\rm imp} = 3$ km/s.
The value of fitting parameter $Q_{\rm RD,0}^{*}$ in Equation~(\ref{eq:fit}) corresponds to the converged $Q_{\rm RD}^{*}$ in the limit of $\Delta x/D_{\rm tar}\rightarrow 0$ (i.e. $n_{\rm tar}\rightarrow\infty$).
From these results, we conclude that ejecta masses and $Q_{\rm RD}^{*}$ obtained by iSALE code also depend on numerical resolution as well as those in SPH simulations by \citet{G15,G17}. 

\begin{figure}[ht]
 \begin{center}
  \includegraphics[width=120mm]{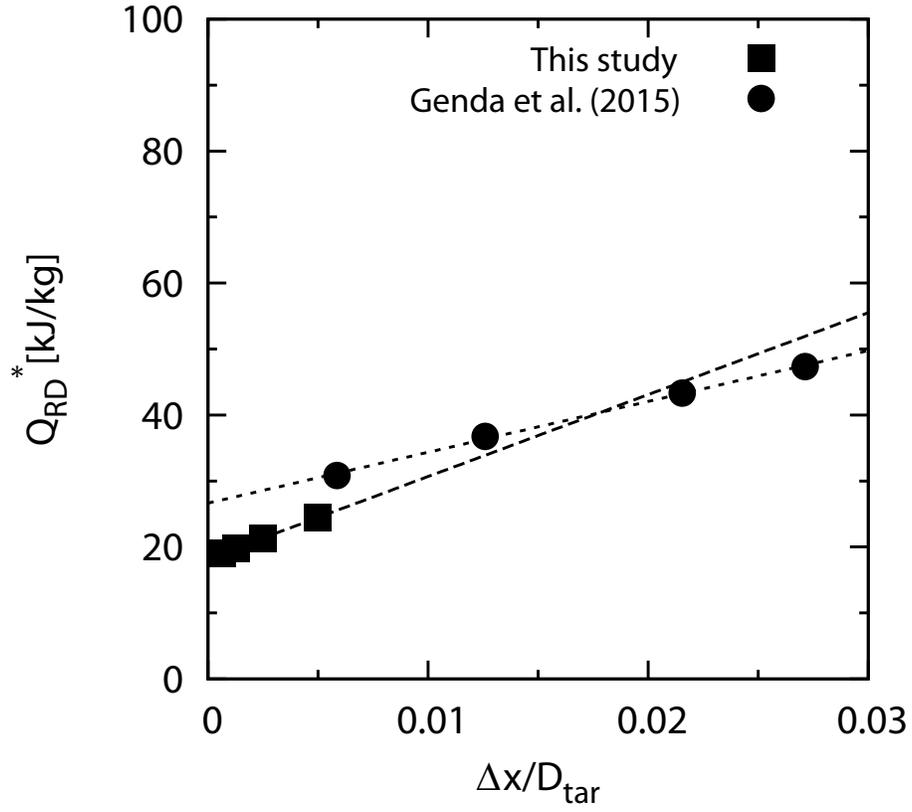}
  \end{center}
 \caption{
Comparison of resolution dependence of $Q_{\rm RD}^{*}$ between this study and \citet{G15}.
The lines are the least-squares fit to the results (Equation~(\ref{eq:fit})).
}
 \label{fig:reso_sph}
\end{figure}

We compare these results with \citet{G15} in detail. 
In the case of $R_{\rm tar}=100$ km and $v_{\rm imp}=3$ km/s,  \citet{G15} carried out SPH simulations for four different numerical resolutions ($N_{\rm SPH}=5\times10^{4}$, $1\times10^{5}$, $5\times10^{5}$, and $5\times10^{6}$, where $N_{\rm SPH}$ is the number of SPH particles in the target) and estimated the converged values of the head-on critical specific impact energy. 
Although their hydrocode is not based on two-dimensional Eulerian code such as iSALE-2D, but three-dimensional Lagrangian code, 
we simply assume that the number of cells on target diameter $2n_{\rm tar}$ is equivalent to $N_{\rm SPH}^{1/3}$
\footnote{When the target consists of $N_{\rm SPH}$ SPH particles, the relationship between $N_{\rm SPH}$ and the number of SPH particles for the target radius $n_{\rm SPH}$ is given as $N_{\rm SPH}^{1/3}=(4\pi/3)^{1/3}n_{\rm SPH}\simeq1.6n_{\rm SPH}$.
However, it is not clear whether $n_{\rm SPH}$ is equal to $n_{\rm tar}$ because of the difference in the numerical scheme.
Thus, we assume $N_{\rm SPH}^{1/3}\simeq1.6n_{\rm SPH}\simeq 2n_{\rm tar}$.}. 
Under the assumption, the case of lowest numerical resolution ($n_{\rm tar}=100$) in this study roughly corresponds to the case of highest resolution ($N_{\rm SPH}=5\times10^{6}$) in \citet{G15}
\footnote{In comparison with SPH code, this correspondence does not change even if we use impactor resolution. 
In \citet{G15}, the impactor is composed of $\sim2\times10^{4}$ SPH particles for the highest resolution case ($N_{\rm SPH}=5\times10^{6}$).
Under the assumption that $2n_{\rm imp}\simeq N_{\rm SPH, imp}^{1/3}$ (where $N_{\rm SPH, imp}$ is the number of SPH particles for the impactor), 
the number of cells per impactor radius $n_{\rm imp}$ is estimated to be 13.6, which is consistent with the value of $n_{\rm imp}$ used by the lowest resolution case (see Table~\ref{table:output}).}. 
Figure~\ref{fig:reso_sph} is the same as Figure~\ref{fig:qd_m100}(b),
but head-on $Q_{\rm RD}^{*}$ obtained from SPH simulations \citep{G15} are also plotted. 
The converged values in both \citet{G15} and in our study are similar values within a 30\% range of each other. 
The slope from numerical data obtained by the iSALE-2D is somewhat steeper than in the case of SPH simulations. 
Thus, for the same numerical resolution, the value of $Q_{\rm RD}^{*}$ estimated by SPH simulations is closer to each of their converged values.

Recently, \citet{G17} found that the dependence of $M_{\rm ej}/M_{\rm tot}$ on $Q_{\rm R}/Q_{\rm RD}^{*}$ is independent of  numerical resolutions: 
once the converged value (or reasonable value of $Q_{\rm RD}^{*}$) is obtained from very high-resolution simulations, 
the general behavior of $M_{\rm ej}/M_{\rm tot}$  is given by low resolution simulations.
Thus, we examine whether the ejected mass due to disruptive collision can be scaled by $Q_{\rm R}/Q_{\rm RD}^{*}$ because they mainly focus on erosive collisions.

\begin{figure}[ht]
 \begin{center}
  \includegraphics[width=120mm]{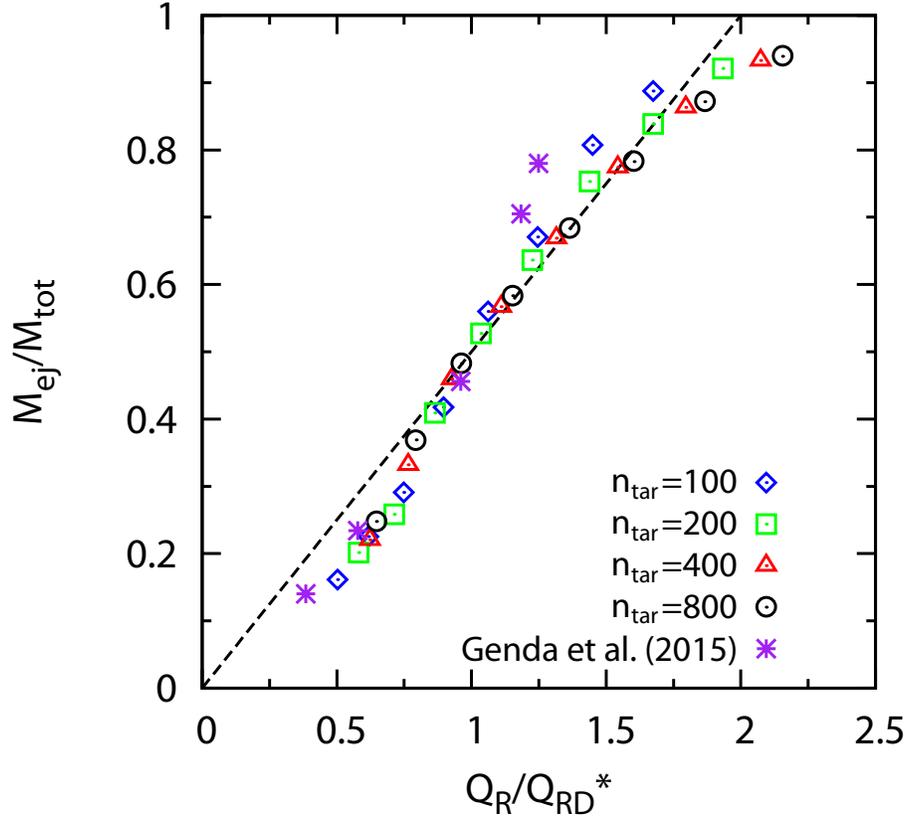}
  \end{center}
 \caption{
Same as Figure~\ref{fig:qd_m100}(a), but $Q_{\rm R}$ is normalized by each $Q_{\rm RD}^{*}$.
Asterisks represent numerical data obtained by \citet{G15}.
Dashed line represents $M_{\rm ej}/M_{\rm tot}$ obtained by the empirical law (Equation~(\ref{eq:uni_ej})).
}
 \label{fig:com_sph}
\end{figure}

Figure~\ref{fig:com_sph} shows the same results as shown in Figure~\ref{fig:qd_m100}(a), but $Q_{\rm R}$ is normalized by each calculated value of $Q_{\rm RD}^{*}$. 
We find that our numerical data can be clearly scaled by $Q_{\rm R}/Q_{\rm RD}^{*}$. 
Asterisks in Figure~\ref{fig:com_sph} represent collision outcomes of  SPH simulations for $N_{\rm SPH}=5\times10^{6}$ \citep{G15}.
Our results agree well with their results despite the different numerical schemes. 
Therefore, this would suggest that even numerical scheme-dependence of ejected mass can be scaled by $Q_{\rm R}/Q_{\rm RD}^{*}$.
\citet{SL09} derived an empirical universal law given by $M_{\rm lrg}/M_{\rm tot}=-0.5(Q_{\rm R}/Q_{\rm RD}^{*}-1)+0.5$ from their numerical simulations.
Using Equation~(\ref{eq:def_mej}), this universal law is written as
\begin{eqnarray}
\frac{M_{\rm ej}}{M_{\rm tot}}=0.5\frac{Q_{\rm R}}{Q_{\rm RD}^{*}}.
\label{eq:uni_ej}
\end{eqnarray}
Equation~(\ref{eq:uni_ej}) is also drawn in Figure \ref{fig:com_sph}.
Our results are in a good agreement with Equation~(\ref{eq:uni_ej}) for $Q_{\rm R}/Q_{\rm RD}^{*}=0.7$ - 1.7.

%%%%%%%%%%%%%%%%%%%%%%%%%%%%%%%%%%%%%%%%

\subsection{Dependence on target size and impact velocity}
\label{subsec:vimp_rtar}
In Section \ref{subsec:reso}, we obtained $Q_{\rm RD}^{*}$ with high accuracy for the case with $R_{\rm tar}=100$ km and $v_{\rm imp}=3$ km/s.
Here, we examine the dependences of $Q_{\rm RD}^{*}$ on the target size and the impact velocity.

\begin{figure}[ht]
 \begin{center}
  \includegraphics[width=120mm]{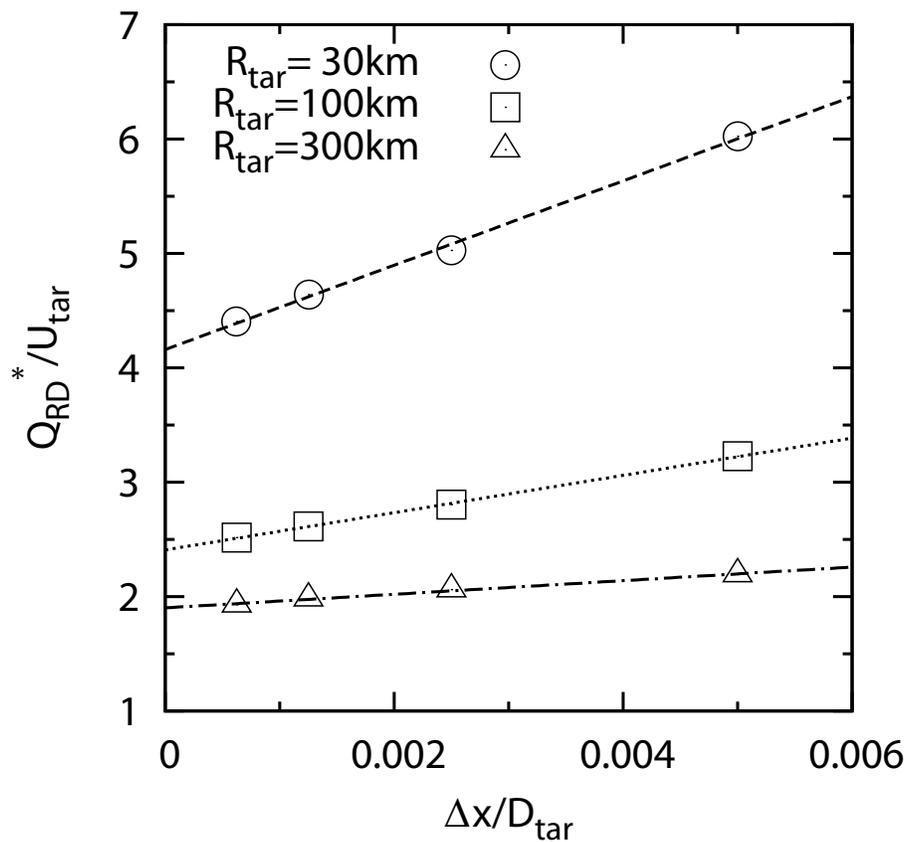}
 \end{center}
 \caption{
 Critical specific impact energy $Q_{\rm RD}^{*}$ normalized by the potential energy of the target $U_{\rm tar}$ as a function of $\Delta x/D_{\rm tar}$ in the case of $v_{\rm imp}=3$ km/s. 
 Circles, squares and triangles represent $R_{\rm tar}=30$, 100 and 300 km, respectively.
 The lines are the least-squares fit to the results based on Equation~(\ref{eq:fit}).
}
 \label{fig:reso_qdpot}
\end{figure}

Figure~\ref{fig:reso_qdpot} shows the dependence of $Q_{\rm RD}^{*}$ on $\Delta x/D_{\rm tar}$  for different target sizes. 
The values of $Q_{\rm RD}^{*}$ for $R_{\rm tar}=30$ and 300 km are estimated by the linear interpolation (Equation (\ref{eq:fit})) in the same way as the case with $R_{\rm tar}=100$ km.
In order to compare the convergence of $Q_{\rm RD}^{*}$ for each target size, $Q_{\rm RD}^{*}$ is normalized by the potential energy at the target's surface $U_{\rm tar}(\equiv GM_{\rm tar}/R_{\rm tar}$, where $G$ is the gravitational constant) \citep[e.g.,][]{M16}, whose values are given by $U_{\rm tar}=0.68$, 7.55, and 67.9 kJ/kg for $R_{\rm tar}=30$, 100, and 300 km, respectively.
Using the least-squares fit to $Q_{\rm RD}^{*}/U_{\rm tar}$ for each target size, 
we estimate the values of $Q_{\rm RD, 0}^{*}/U_{\rm tar}$ for $R_{\rm tar}=30$ and 300 km are 4.16 and 1.90, which correspond to $Q_{\rm RD, 0}^{*}=2.83$ and 129.01 kJ/kg, respectively. 
Therefore, $Q_{\rm RD, 0}^{*}/U_{\rm tar}$ is proportional to $R_{\rm tar}^{-0.34}$, which leads to $Q_{\rm RD,0}^{*}\propto R_{\rm tar}^{1.66}$ for constant density and impact velocity.
We also find that the slopes of the lines  become steeper in the case of smaller target size \citep[see also][]{G17}.
This is explained by low resolutions of impactors for small targets 
because small mass ratios between impactors and targets are required to find $Q_{\rm RD}^{*}$ for small targets.
This result also suggests that the value of $Q_{\rm RD}^{*}$ for a large target is closer to the converged value in the case where the resolution is the same.  

\begin{figure}[ht]
 \begin{center}
  \includegraphics[width=120mm]{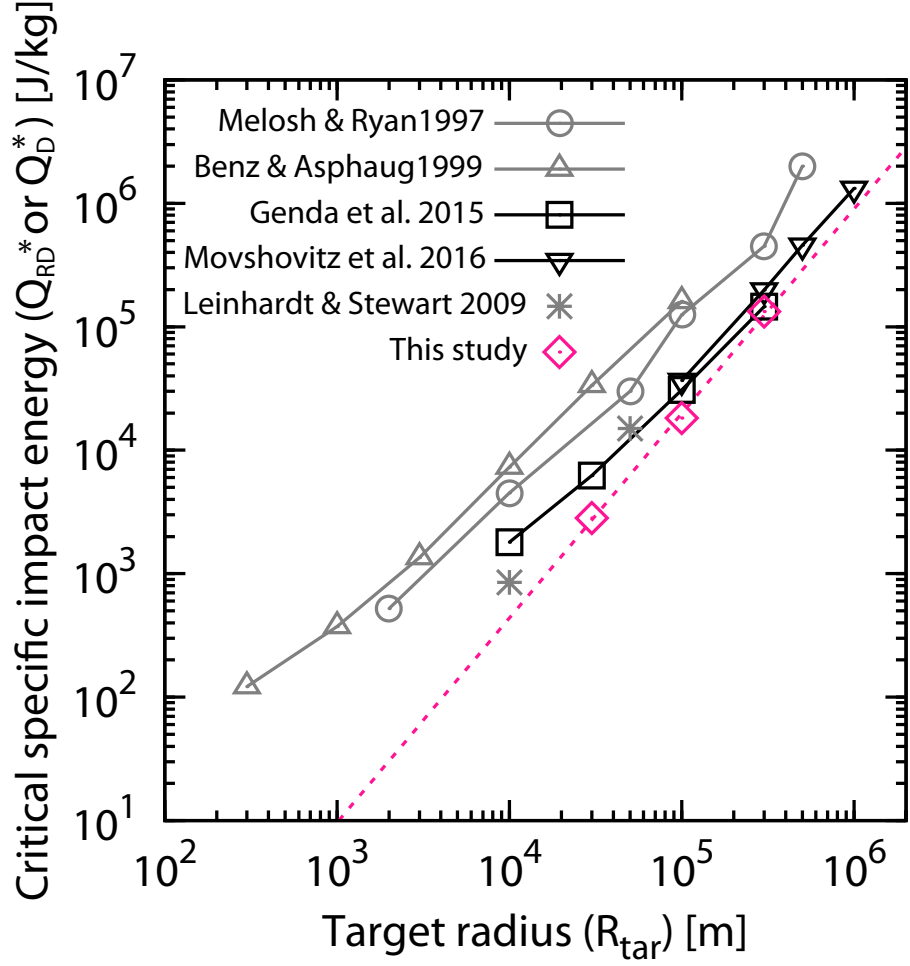}
  \end{center}
 \caption{Dependence of $Q_{\rm RD}^{*}$ obtained by head-on impact simulations as a function of target radius. 
The diamond symbols represent our results, which  are converged values ($Q_{\rm RD, 0}^{*}$) for the case with 3 km/s.
Other symbols are the values of critical specific impact energy ($Q_{\rm RD}^{*}$ or classical definition of $Q_{\rm D}^{*}(=0.5M_{\rm imp}v_{\rm imp}^{2}/M_{\rm tot}$)) obtained by various previous studies.
Gray data symbols represent cases for planetesimals with material strength and/or damage. 
Black symbols represent purely hydrodynamic cases.
The dashed line is the least-squares fit to our results.
 }
 \label{fig:rtar_qd}
\end{figure}

Figure~\ref{fig:rtar_qd} shows the dependence of the critical specific impact energy for head-on collision on target sizes.
In Figure~\ref{fig:rtar_qd}, the diamond symbols represent $Q_{\rm RD, 0}^{*}$  obtained from the present work and the other symbols are the head-on critical specific impact energy from previous works.
The critical specific impact energy in the gravity regime is known to increase with $R_{\rm tar}$.
In fact, $Q_{\rm RD, 0}^{*}$ estimated by the present work also increases as target radius increases.
We find that the values of $Q_{\rm RD, 0}^{*}$  become rather low compared to the critical specific impact energy obtained by some previous works that consider material strength and/or damage. 
On the other hand, our results are roughly consistent with the values of $Q_{\rm RD}^{*}$ obtained by a purely hydrodynamic target \citep{G15, M16}, although their numerical scheme is different from ours.
Thus, it seems that the dependence of the values of the critical specific impact energy on numerical methods becomes small in the case of high numerical resolution.

\begin{figure}[ht]
 \begin{center}
  \includegraphics[width=120mm]{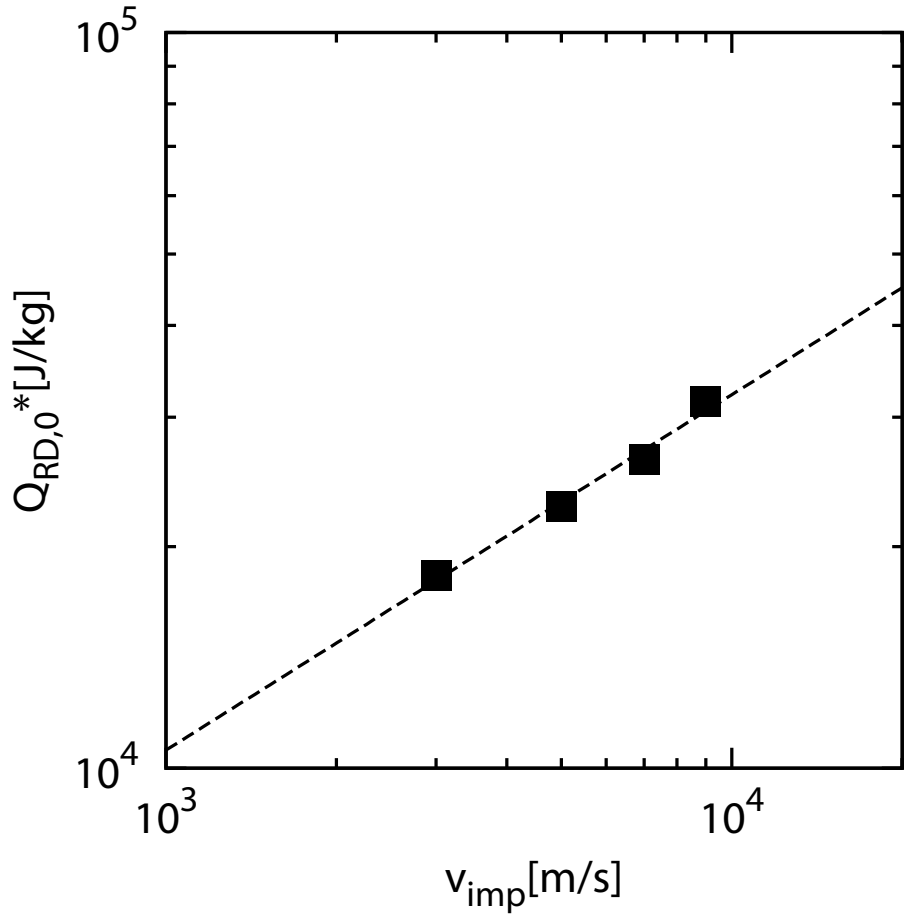}
 \end{center}
 \caption{
 Dependence of converged values $Q_{\rm RD,0}^{*}$ on impact velocity ($v_{\rm imp}=3$, 5, 7, and 9 km/s) in the case of $R_{\rm tar}=100$ km.
The dashed line is the least-squares fit to the results.
}
 \label{fig:vimp_qrd}
\end{figure}

Figure~\ref{fig:vimp_qrd} shows the dependence of $Q_{\rm RD,0}^{*}$ on impact velocity $v_{\rm imp}$ ($v_{\rm imp}=3$, 5, 7, and 9 km/s). 
In order to examine impact velocity dependence, the target size is fixed at 100 km. 
The values of $Q_{\rm RD,0}^{*}$ somewhat depend on impact velocities in the range of $v_{\rm imp}=3$ - 9 km/s. 
When the impact velocities are rather high, impact energies would be easily converted into internal energy \citep{HH90}. 
As a result, with increasing impact velocity, the kinetic energy of ejecta decreases, which leads to the disruption of the target being hindered. 
From our numerical results, $Q_{\rm RD,0}^{*}$ is proportional to $v_{\rm imp}^{0.48}$.

Here, we compare our results with the scaling law.
In the case of the gravity regime, $M_{\rm lrg}/M_{\rm tot}$ is described by a function of the scaling parameter \citep{HH90}, 
\begin{eqnarray}
\Pi_{\rm G}=(\rho G)^{-3\mu/2}R_{\rm tar}^{-3\mu}v_{\rm imp}^{3\mu-2}Q_{\rm R},
\label{eq:scal_pi}
\end{eqnarray}
where $\rho$ is the density of the target,  
and $\mu$ is a constant value dependent on material (Table~\ref{table:qg}).
Therefore, 
\begin{eqnarray}
\frac{M_{\rm lrg}}{M_{\rm tot}}=F(\Pi_{\rm G}),
\label{eq:scal_mlar}
\end{eqnarray}
where $F(\Pi_{\rm G})$ is a monotonically increasing function of $\Pi_{\rm G}$. 
For $M_{\rm lrg}/M_{\rm tot}=0.5$, $\Pi_{\rm G}$ is equal to a constant value $\Pi_{\rm G}^{*}$ and also $Q_{\rm R}=Q_{\rm RD}^{*}$.
Hence, from Equation~(\ref{eq:scal_pi}), $Q_{\rm RD}^{*}$ can be given as \citep{LS12, M16},
\begin{eqnarray}
Q_{\rm RD}^{*}=\Pi_{\rm G}^{*}(\rho G)^{3\mu/2}R_{\rm tar}^{3\mu}v_{\rm imp}^{2-3\mu}.
\label{eq:scal_pistar}
\end{eqnarray}
Using Equations (\ref{eq:scal_pi}) and (\ref{eq:scal_pistar}), we have
\begin{eqnarray}
\Pi_{\rm G}=\Pi_{\rm G}^{*}\frac{Q_{\rm R}}{Q_{\rm RD}^{*}}.
\label{eq:pigpig}
\end{eqnarray}

From Equation~(\ref{eq:scal_pistar}), the dependence of $Q_{\rm RD}^{*}$ on $R_{\rm tar}$ and $v_{\rm imp}$ can be approximated by a power-law given by 
\begin{eqnarray}     
Q_{\rm RD}^{*}\propto R_{\rm tar}^{p}v_{\rm imp}^{q}
\label{eq:pow}
\end{eqnarray}
where $p$ and $q$ are fitting parameters. 
Based on our numerical results shown in Figures~\ref{fig:rtar_qd} and \ref{fig:vimp_qrd}, 
the dependence becomes $p=1.66$ and $q=0.48$. 
According to the scaling law,  the values of $p$ and $q$ depend on the value of $\mu$ (see Equation~(\ref{eq:scal_pistar})).
Then, from Figure~\ref{fig:rtar_qd}, we can estimate the value of $\mu$ to be 0.55, 
which is in excellent agreement with the value of $\mu$ if the target is composed of non-porous material (e.g. water and rock) (see Table~\ref{table:qg}). 
However, the value of $\mu$ obtained by Figure~\ref{fig:vimp_qrd} becomes smaller than the case with non-porous material.
On the other hand, \citet{MR97} analytically examined the dependence of $Q_{\rm RD}^{*}$ on $R_{\rm tar}$ and $v_{\rm imp}$, and they obtained $p=1.5$ and $q=0.5$. 
Thus, the target size dependence agrees  well with the scaling law, while the velocity dependence is consistent with analytic consideration.

%%%%%%%%%%%%%%%%%%%%%%%%%%%%%%%%%%%%%%%%%%%%%%%%%%%%%%%%%%%%%%%%%
 %%%%%%%%%%%%%%%%%%%%%%%%%%%%%%%%%%%%%%%%%%%%%%%%%%%%%%%%%%%%%%%%%
\section{CONNECTION BETWEEN EROSIVE AND DISRUPTIVE COLLISIONS}
\label{sec:result2}

In this section, we examine $Q_{\rm R}$-dependence of the ejecta mass for a wide range of impactor radii $R_{\rm imp}$, including erosive collisions.  
The numerical methods are basically the same as those described in Section \ref{sec:num}, but slightly modified. 
In the previous sections, the number of cells for the target $n_{\rm tar}$ was fixed because we only consider disruptive collisions in which the target is entirely and largely deformed. 
However, in the case of erosive collision, large deformation due to the impact appears near the impact point and its area depends on impactor size. 
Thus, in this section, we fixed the number of cells for the impactor $n_{\rm imp}$ (see Figure~\ref{fig:illust}), while the value of $n_{\rm tar}$ is changed depending on the impactor size. 
Typically, we set $n_{\rm imp}=20$.

\begin{figure}[ht]
 \begin{center}
 \includegraphics[width=75mm]{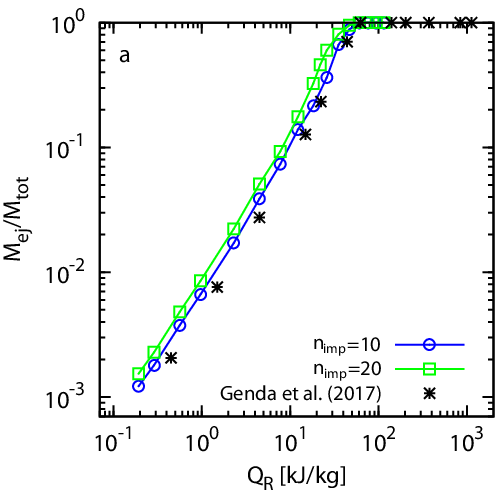}
 \includegraphics[width=75mm]{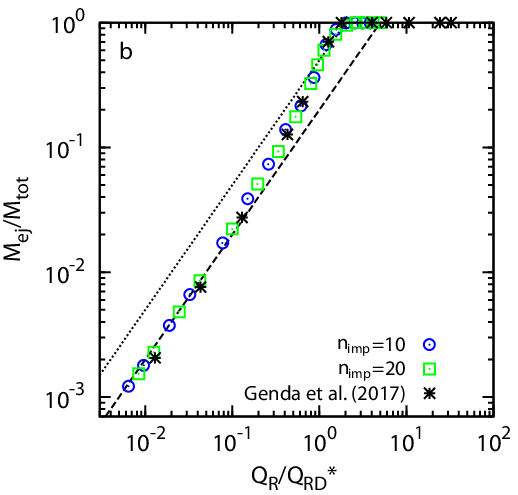}
 \end{center}
 \caption{(a) Dependence of normalized mass of ejecta on numerical resolution in the case of $R_{\rm tar}=100$ km and $v_{\rm imp}=3$ km/s.
Circles and squares represent outcomes of $n_{\rm imp}=10$ and 20, respectively.
Asterisks represent numerical data obtained by \citet{G17}. 
Panel (b) is the same as panel (a), but $Q_{\rm R}$ is normalized by each $Q_{\rm RD}^{*}$.
Dashed and dotted lines represent $M_{\rm ej}/M_{\rm tot}$ obtained by Equation~(\ref{eq:spara}) with $\psi=0.4$ and the empirical law (Equation~(\ref{eq:uni_ej}) or Equation~(\ref{eq:spara}) with $\psi=1$), respectively.
}
 \label{fig:reso_ero}
\end{figure}

First, we examine the dependence of erosive collisions on numerical resolution in the same way as we did in Section~\ref{sec:result1}.
Figure~\ref{fig:reso_ero}(a) shows that mass ejected by erosive collisions in the case of $R_{\rm tar}=100$ km and $v_{\rm imp}=3$ km/s as a function of $Q_{\rm R}$.  
Blue circles and green squares represent collision outcomes for $n_{\rm imp}=10$, and 20, respectively. 
We find that the masses ejected by erosive collisions also depend on numerical resolution and become larger due to the higher numerical resolution. 
Figure~\ref{fig:reso_ero}(b)  shows ejected mass as a function of $Q_{\rm R}$ normalized by each calculated value of $Q_{\rm RD}^{*}$. 
We make new estimates for the values of $Q_{\rm RD}^{*}$ based on numerical data in Figure~\ref{fig:reso_ero}(a); $Q_{\rm RD}^{*}=30.0$ and 22.9 kJ/kg  for $n_{\rm imp}=10$ and 20, respectively. 
Lower resolution simulations result in larger $Q_{\rm RD}^{*}$. 
However, we find that $M_{\rm ej}/M_{\rm tot}$ is nicely scaled by $Q_{\rm R}/Q_{\rm RD}^{*}$ (Figure~\ref{fig:reso_ero}(b)). 
Asterisks in Figure~\ref{fig:reso_ero}(b) represent numerical results obtained by SPH impact simulations with the same impact conditions \citep {G17}. 
We find that their features agree very well with our results.  
However, we also note that the material properties of planetesimals are neglected in both simulations.
Since the material properties affect collisional outcomes (see Section~\ref{sec:dis}), 
it is also necessary to compare outcomes between different impact simulations including material properties,
which is out of the scope of this study.

\begin{figure}[ht]
 \begin{center}
   \includegraphics[width=120mm]{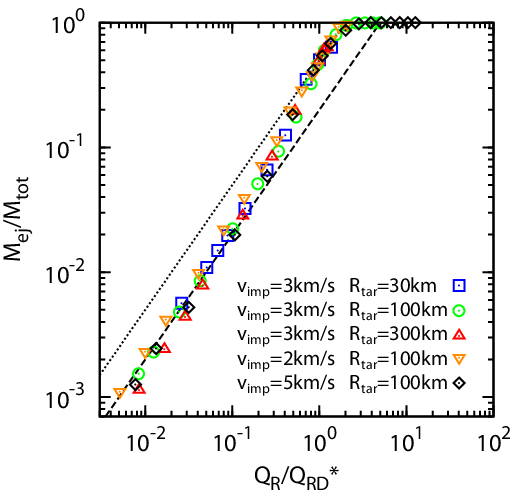}
 \end{center}
 \caption{
 Dependence of $M_{\rm ej}/M_{\rm tot}$ on $Q_{\rm R}/Q_{\rm RD}^{*}$ for five different impact conditions in the case of $n_{\rm imp}=20$. 
Squares, circles, triangles,  inverted triangles, and diamonds represent the results for ($v_{\rm imp}$, $R_{\rm tar}$)=(3 km/s, 30 km), (3 km/s, 100 km), (3 km/s, 300 km), (2 km/s, 100 km), and (5 km/s, 100 km), respectively.
Dashed and dotted lines represent $M_{\rm ej}/M_{\rm tot}$ obtained by Equation~(\ref{eq:spara}) with $\psi=0.4$ and the empirical law (Equation~(\ref{eq:uni_ej}) or Equation~(\ref{eq:spara}) with $\psi=1$), respectively.
}
 \label{fig:simi_ero}
\end{figure}

Next, we examine the dependence of erosive collisions on target size and impact velocity.
Figure~\ref{fig:simi_ero} shows the dependence of the ejected mass for five different cases of impact conditions 
as a function of impact energy normalized by each calculated value of $Q_{\rm RD}^{*}$.
Although the slope of numerical data for the case with low impact velocity ($v_{\rm imp}=2$ km/s) is slightly different from the others, 
$M_{\rm ej}/M_{\rm tot}$ is clearly scaled by $Q_{\rm R}/Q_{\rm RD}^{*}$. 
We also confirm that our results are consistent with the dependence on $R_{\rm tar}$ and $v_{\rm imp}$ obtained by \citet{G17}.
From Equation~(\ref{eq:pigpig}), the ratio $Q_{\rm R}/Q_{\rm RD}^{*}$ is equal to $\Pi_{\rm G}/\Pi_{\rm G}^{*}$.
Thus we can say that the scaling in Figure~\ref{fig:simi_ero} is equivalent to the scaling law of $M_{\rm lrg}(=M_{\rm tot}-M_{\rm ej})$ for the gravity regime proposed by \citet{HH90}.
For erosive collisions with $Q_{\rm R}/Q_{\rm RD}^{*}\ll1$, 
the obtained ejected mass is fitted by a linear relation,
\begin{eqnarray}
\frac{M_{\rm ej}}{M_{\rm tot}}=\frac{\psi}{2}\left(\frac{Q_{\rm R}}{Q_{\rm RD}^{*}}\right). 
\label{eq:spara}
\end{eqnarray}
From Figures \ref{fig:reso_ero}(b) and \ref{fig:simi_ero}, the non-dimensional parameter $\psi$ is obtained as 0.4, 
independent of the numerical resolution, the target size, and the impact velocity.
The value of $\psi$ depends on the impact angle and becomes larger with oblique collisions.
According to SPH simulations by \citet{G17}, $\psi$ is 1.2 for the typical oblique (45$^\circ$) collision.

The ejected masses from erosive collisions are also described by the crater scaling law \citep{H93, HH11}.
When the densities of the target and the impactor are the same, the total mass of fragments with velocity greater than $v$ can be given by \citep{HH11}
\begin{eqnarray}
\frac{M(v)}{M_{\rm imp}}=\frac{3k}{4\pi}C_{0}^{3\mu}\left( \frac{v}{v_{\rm imp}}\right)^{-3\mu},
\label{eq:ctr}
\end{eqnarray}
where $k$ and $C_{0}$ are constants whose values are dependent on target material (Table~\ref{table:qg}).
A fragment with a velocity higher than the escape velocity  of the target $v_{\rm esc}(\equiv\sqrt{2GM_{\rm}/R_{\rm tar}})$ is not bound by the target's gravity.
Therefore, $M(v_{\rm esc})$ would correspond to the ejected mass $M_{\rm ej}$ obtained in this study. 
In the case of erosive collisions, since the mass of the impactor is significantly lower than that of the target, the specific impact energy is given as (i.e. classical definition of  specific impact energy) 
\begin{eqnarray}
Q_{\rm R}\simeq 0.5M_{\rm imp}v_{\rm imp}^{2}/M_{\rm tot}.
\label{eq:q_cla}
\end{eqnarray}
Using Equations (\ref{eq:q_cla}) and (\ref{eq:ctr}) with $M_{\rm ej}=M(v_{\rm esc})$, $M_{\rm ej}$ can be written as
\begin{eqnarray}
\frac{M_{\rm ej}}{M_{\rm tot}}=\frac{3k}{2\pi}C_{0}^{3\mu}\left( \frac{v_{\rm imp}}{v_{\rm esc}}\right)^{3\mu-2}\left(\frac{Q_{\rm R}}{v_{\rm esc}^{2}}\right).
\label{eq:mej_ctr}
\end{eqnarray}

Equation~(\ref{eq:mej_ctr}) is essentially the crater scaling law. 
Thus we assume that the value of $Q_{\rm R}$ is considerably smaller than $Q_{\rm RD}^{*}$. 

Then, since Equation~(\ref{eq:mej_ctr}) should be equal to Equation~(\ref{eq:spara}), we obtain the following semi-analytic formula for $Q_{\rm RD}^{*}$ as
\begin{eqnarray}
Q_{\rm RD}^{*}=\frac{\psi\pi}{3k}C_{0}^{-3\mu}v_{\rm esc}^{2}\left(\frac{v_{\rm imp}}{v_{\rm esc}}\right)^{2-3\mu},
\label{eq:sscale_v}
\end{eqnarray}
or
\begin{eqnarray}
Q_{\rm RD}^{*}=\frac{\psi}{8k}\left(\frac{8\pi}{3}\right)^{(3\mu +2)/2}C_{0}^{-3\mu}\left(\rho G\right)^{3\mu /2}R_{\rm tar}^{3\mu}v_{\rm imp}^{2-3\mu}.
\label{eq:sscale}
\end{eqnarray}
From Equation~(\ref{eq:scal_pistar}),  $\Pi_{\rm G}^{*}$ is written by
\begin{eqnarray}
\Pi_{\rm G}^{*}=\frac{\psi}{8k}\left(\frac{8\pi}{3}\right) ^{(3\mu +2)/2}C_{0}^{-3\mu}.
\label{eq:s_pi}
\end{eqnarray}
Although Equation~(\ref{eq:sscale}) has the same functional form derived by previous studies \citep[e.g.][]{HH90, LS12, M16}, 
the difference is that, except for $\psi$, the value of $\Pi_{\rm G}^{*}$ is determined not by numerical data, but by experimental data.
For example, in the case of rock (see, Table \ref{table:qg}), $Q_{\rm RD}^{*}$ is written as, 
\begin{eqnarray}
Q_{\rm RD}^{*}=3.6\times10^{4}\left(\frac{R_{\rm tar}}{100 \;{\rm km}}\right)^{1.65}\left(\frac{v_{\rm imp}}{3\; {\rm km/s}}\right)^{0.35} \;\;\;{\rm J/kg},
\label{eq:qrd_rock}
\end{eqnarray}
where we assume $\psi=0.4$.

\begin{figure}[ht]
 \begin{center}
  \includegraphics[width=120mm]{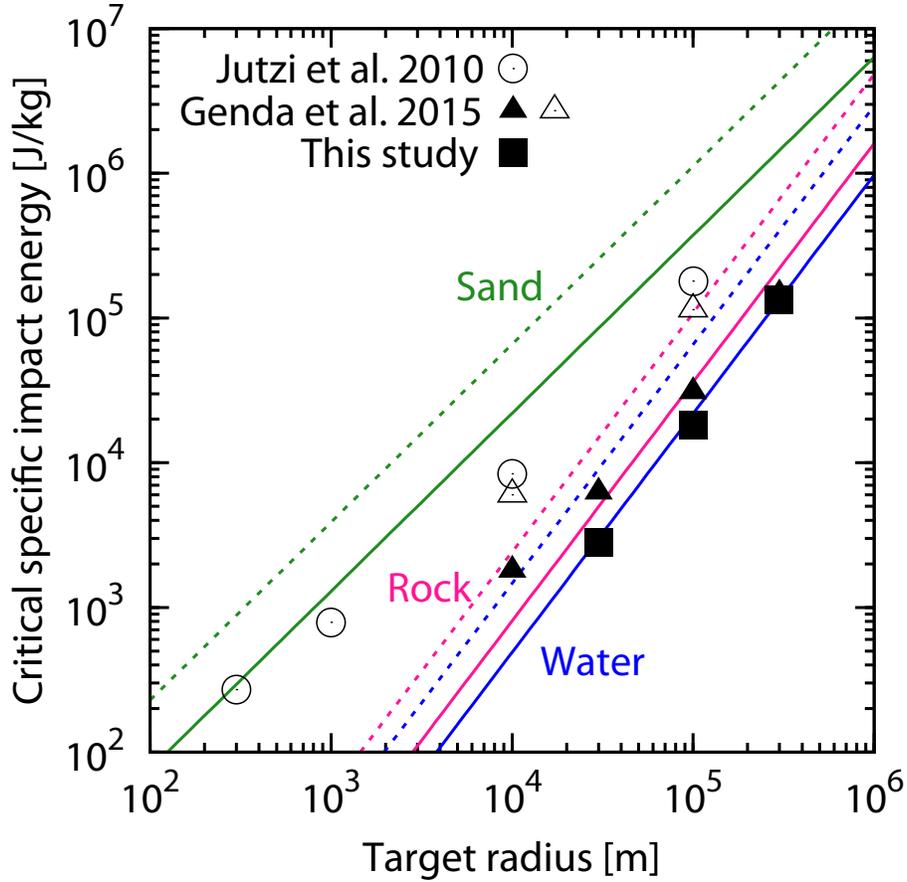}
 \end{center}
 \caption{Comparison of $Q_{\rm RD}^{*}$ between semi-analytic formula and numerical results.
Blue, red, and green lines correspond to $Q_{\rm RD}^{*}$ for the case with water, rock and sand, respectively.
Solid and dotted lines represent head-on collisions ($\psi=0.4$) and oblique (45$^\circ$) collisions ($\psi=1.2$), respectively.
Points show $Q_{\rm RD}^{*}$ for three different numerical data.
Filled and open symbols represent head-on collisions and oblique (45$^\circ$) collisions, respectively.
}
 \label{fig:psi_scale}
\end{figure}

Figure~\ref{fig:psi_scale} shows the dependence of $Q_{\rm RD}^{*}$  on target size obtained by Equation~(\ref{eq:sscale}) for several materials. 
In the case of non-porous material (water and rock), the values of $Q_{\rm RD}^{*}$ for rock are larger than they are in the case with water. 
This is because the gravitational potential increases as the density increases. 
In comparison with the critical specific values obtained by iSALE code and SPH code, 
the values of $Q_{\rm RD}^{*}$ for the case with rock are roughly consistent with the numerical results.
In the case of oblique collision we set $\psi=1.2$, but  we assume that material parameters are unchanged 
because it is generally thought that the crater size hardly depends on impact angle, except for in the case of a very low impact angle \citep{M11}.
The value of $Q_{\rm RD}^{*}$ at $R_{\rm tar}=100$ km agrees with the numerical data obtained by \citet{G15},
while deviation between semi-analytic and numerical results for $R_{\rm tar}=10$ km becomes large.

For sand, the slope becomes smaller compared with non-porous material because the value of $\mu$ depends on the porosity of the target, and it decreases with increasing degrees of porosity (Table~\ref{table:qg}). 
Also, energy dissipation by compaction takes place within porous targets such as sand.
Thus, the value of $Q_{\rm RD}^{*}$ for sand is much higher than it is for the case with non-porous material. 
The circles in Figure~\ref{fig:psi_scale} represent numerical results obtained by \citet{J10}, who performed oblique impact simulation using the SPH method, including the effect of porosity. 
Indeed, their numerical data also reflects the effect of the porosity: a small slope and large critical specific impact energy.
However, we find that semi-analytic results are significantly different from \citet{J10}.
Since the dependence of material parameters on the impact angle is small,  
the deviation between semi-analytic and numerical result would be caused by the value of $\psi$. 
Therefore, further studies are needed to clarify the effect of material properties on non-dimensional parameter $\psi$. 

%%%%%%%%%%%%%%%%%%%%%%%%%%%%%%%%%%%%%%%%%%%%%%%%%%%%%%%%%%%%%%%%%%%%%%
%%%%%%%%%%%%%%%%%%%%%%%%%%%%%%%%%%%%%%%%%%%%%%%%%%%%%%%%%%%%%%%%%%%%%%
\section{\bf DISCUSSION}
\label{sec:dis}
\subsection{\bf Effect of oblique impacts}
So far, we have focused on the case of head-on collisions with the impact angle $\theta=0^\circ$.
Here, we discuss the effect of oblique impacts on our results.

In the previous section, we have obtained the linear relation between ejecta mass and impact energy (Equation~(\ref{eq:spara})).
Even if ejected mass is averaged over impact angles, the linear relation holds: $\bar{M}_{\rm ej}/M_{\rm tot}\propto Q_{\rm R}/\bar{Q}_{\rm RD}^{*}$ where overlines  denote angle-averaged quantities \citep{G17}. 
Thus, the linear relation for the oblique impact case is written by
\begin{eqnarray}
\frac{\bar{M}_{\rm ej}}{M_{\rm tot}}=\frac{\bar{\psi}}{2}\frac{Q_{\rm R}}{\bar{Q}_{\rm RD}^{*}},
\label{eq:ave_sim}
\end{eqnarray}
where $\bar{\psi}$ is angle-averaged $\psi$ whose value is 0.88 for a purely hydrodynamic case \citep[see][]{G17}.

As we have shown, in the framework of the crater scaling law,  the total mass of fragments with velocity greater than $v$ can be scaled by $v/v_{\rm imp}$ (Equation~(\ref{eq:ctr})).
On the other hand,  although oblique impact experiments have not been sufficiently performed, it is thought that the total mass of fragments formed by an oblique impact $M(v,\theta)$ can be scaled by the normal component of the impact velocity, $v_{\rm imp}\cos\theta$  \citep{HH11}.
Under this approximation, the crater scaling law for the oblique impact is  given by
\begin{eqnarray}
\frac{M(v,\theta)}{M_{\rm imp}}=\frac{3k}{4\pi}C_{0}^{3\mu}\left( \frac{v}{v_{\rm imp}\cos\theta}\right)^{-3\mu}.
\label{eq:ang_v_mej}
\end{eqnarray}
Since  the total ejecta mass at an oblique impact with the angle $\theta$, $M_{\rm ej}(\theta)$, is equal to $M({v}_{\rm esc}, \theta )$, the total ejecta mass $M_{\rm ej}(\theta)$ is obtained from Equation~(\ref{eq:ang_v_mej}) as
\begin{eqnarray}
\frac{M_{\rm ej}(\theta)}{M_{\rm imp}}=\frac{M_{\rm ej}(0^\circ)}{M_{\rm imp}}\left(\cos\theta\right)^{3\mu}.
\label{eq:ang_mej}
\end{eqnarray}
%where $M_{\rm ej}(\theta)$ is ejected mass of each impact angle.
Using the probability distribution for impact angle \citep{S62}, angle-averaged ejected mass $\bar{M_{\rm ej}}/M_{\rm imp}$ is given by \citep[see also][]{G17}
\begin{eqnarray}
\frac{\bar{M_{\rm ej}}}{M_{\rm imp}}&=&\int_{0^\circ}^{90^\circ} 2 \frac{M_{\rm ej}(0^\circ)}{M_{\rm imp}}\sin\theta\left(\cos\theta \right)^{3\mu+1} d\theta \nonumber \\
                                                              &=& \frac{2}{2+3\mu}\frac{M_{\rm ej}(0^\circ)}{M_{\rm imp}}.
\label{eq:ave_mej}                                                              
\end{eqnarray}
The amount of mass ejected by head-on erosive collision is reduced by the factor $2/(2+3\mu)$ due to the effect of impact angle.
In the case of a purely hydrodynamic body, the factor becomes 0.548, which is consistent with numerical results obtained by \citet{G17}.

Using Equations~(\ref{eq:mej_ctr}), (\ref{eq:ave_sim}), and (\ref{eq:ave_mej}), $\bar{Q}_{\rm RD}^{*}$ can be given by,
\begin{eqnarray}
\bar{Q}_{\rm RD}^{*}=\frac{2+3\mu}{2}\frac{\bar{\psi}}{8k}\left(\frac{8\pi}{3}\right)^{(3\mu +2)/2}C_{0}^{-3\mu}\left(\rho G\right)^{3\mu /2}R_{\rm tar}^{3\mu}v_{\rm imp}^{2-3\mu}.
\end{eqnarray}
$\bar{Q}_{\rm RD}^{*}$ for a rocky planetesimal without material strength is written as,
\begin{eqnarray}
\bar{Q}_{\rm RD}^{*}=1.4\times10^{5}\left(\frac{R_{\rm tar}}{100 \;{\rm km}}\right)^{1.65}\left(\frac{v_{\rm imp}}{3\; {\rm km/s}}\right)^{0.35} \;\;\;{\rm J/kg}.
\label{eq:ave_rock}
\end{eqnarray}
Equation~(\ref{eq:ave_rock}) shows that the value of $\bar{Q}_{\rm RD}^{*}$ is four times as large as that of head-on $Q_{\rm RD}^{*}$ (see Equation~(\ref{eq:qrd_rock})).
Indeed, \citet{G17} examined  the dependence of $Q_{\rm RD}^{*}$ for the case of a purely hydrodynamic body on impact angle, and estimated the value of $\bar{Q}_{\rm RD}^{*}$ to be $1.81 \times 10^{5}$ J/kg, which is close to the value of $\bar{Q}_{\rm RD}^{*}$ obtained by the above semi-analytic formula (Equation~(\ref{eq:ave_rock})).

\subsection{\bf Effects of material properties}
In this study, we had assumed that target and impactor planetesimals are purely hydrodynamic bodies including self-gravity, 
but they would  have material strength, friction and compaction.
Recently, \citet{J15} examined the dependence of collision outcomes on material properties of the target, 
and showed that the mass of  the largest body after disruptive collisions becomes considerably large compared to the case with a purely hydrodynamic body.
Although the effect of the material properties has not been taken into account in this study so far, 
the iSALE-2D can deal with several energy dissipation mechanisms inside planetary bodies.

\begin{figure}[ht]
 \begin{center}
  \includegraphics[width=120mm]{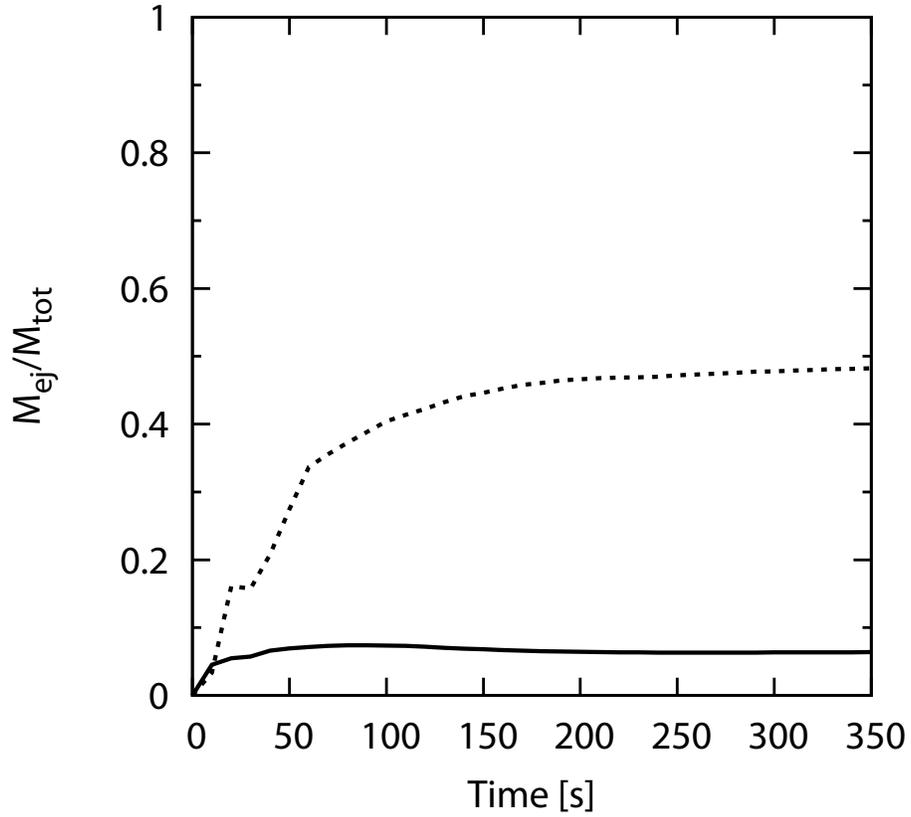}
 \end{center}
 \caption{Comparison of $M_{\rm ej}/M_{\rm tot}$ between the purely hydrodynamic case and the case with material properties. 
Except for the setting for material properties, impact conditions in Figure~\ref{fig:snapshot} are used.
Solid and dotted  curves represent the case with material strength and damage (see also Table~\ref{table:material}), 
and the case of a purely hydrodynamic body shown in Figure~\ref{fig:snapshot}, respectively.
}
 \label{fig:dame}
\end{figure}

As a demonstration of this effect, we further perform a simulation of collisions between planetesimals including the effects of material strength and damage; 
the parameter values are listed in Table~\ref{table:material}.
Figure~\ref{fig:dame} shows the difference in $M_{\rm ej}/M_{\rm tot}$ between  the purely hydrodynamic case and the case with material strength and damage.  
We use the same impact conditions as in Figure~\ref{fig:snapshot}, except for the setting for material properties of planetesimals.
Although the impact energy is unchanged, the ejected mass becomes significantly smaller due to energy dissipation.
The value of $M_{\rm ej}/M_{\rm tot}$ eventually converges to $M_{\rm ej}\simeq0.07M_{\rm tot}$ at $t=350$ s.
This result means  that the value of $Q_{\rm RD}^{*}$ substantially increases due to material strength and damage \citep{J15}.
In addition to such effects, the effect of porosity also plays an important role for the determination of $Q_{\rm RD}^{*}$.
It is generally thought that the value of $Q_{\rm RD}^{*}$ becomes large via impact energy dissipation due to compaction \citep{J15}.
However, since the effect of porosity depends on the density and the strength of the target,  $Q_{\rm RD}^{*}$ would become small due to inefficient reaccumulation of a porous target after collision \citep{J10}.
Therefore,  we will investigate the effects of material properties on $Q_{\rm RD}^{*}$ in the future.

%%%%%%%%%%%%%%%%%%%%%%%%%%%%%%%%%%%%%%%%%%%%%%%%%%%%%%%%%%%%%%%%%%%%%%

\section{SUMMARY}
\label{sec:sum}
In this study, we have performed head-on impact simulations of purely hydrodynamic planetesimals in the gravity regime by using shock-physics code iSALE-2D,
and have made a comparison of collisional outcomes between the SPH code and iSALE-2D code.
We found our numerical simulation results agree well with those obtained by the SPH simulation.
Detailed analysis gives the following three results.
\begin{itemize}
\item The value of $Q_{\rm RD}^{*}$ depends on numerical resolution. 
With decreasing the spatial cell size $\Delta x$, the differences in the values of $Q_{\rm RD}^{*}$ between numerical resolutions linearly decrease and would be close to convergence. Thus, the converged value of $Q_{\rm RD}^{*}$ at $\Delta x \rightarrow 0$ can be estimated by the least-squares fit to $Q_{\rm RD}^{*}$ for each numerical resolution.
\item The converged value $Q_{\rm RD,0}^{*}$ obtained by the iSALE code is similar to the case of the SPH code, and the difference in $Q_{\rm RD,0}^{*}$ between them is within a 30\% range of variation.
\item The relationship between ejected mass normalized by total mass ($M_{\rm ej}/M_{\rm tot}$) and impact energy $Q_{\rm R}$ generally depends on various impact conditions. However, when $Q_{\rm R}$ is scaled by $Q_{\rm RD}^{*}$ that is calculated for each impact simulation, 
the relationship is independent of numerical resolution, impact velocity and target size. 
This similarity law for $Q_{\rm R}/Q_{\rm RD}^{*}$ is confirmed for a wide range of specific impact energy from disruptive collisions to erosive collisions. 
\end{itemize}
Using the above similarity law and the crater scaling law, 
we obtained a semi-analytic formula for the critical specific impact energy $Q_{\rm RD}^{*}$ (Equation~(\ref{eq:sscale})). 
In the case of a non-porous object, the values of $Q_{\rm RD}^{*}$ estimated by the semi-analytic formula agree with  numerical results obtained by the iSALE code and SPH code. 
However, the values of $Q_{\rm RD}^{*}$ for porous objects are  inconsistent with numerical results from SPH simulation taking into account  the effect of porosity \citep{J10}.
Thus, the value of $\psi$ would depend on material properties, as the value of $\mu$ depends on the porosity of the target.

As mentioned above, most of our results can reproduce the results obtained from SPH simulations by \citet{G15, G17} despite different numerical methods. 
Especially, the correspondence of $Q_{\rm RD}^{*}$ would help us better understand planet formation. 
\citet{K10} assumed a simple fragmentation model describing both disruptive collision and erosive collision, and analytically derived the final mass of protoplanets formed in the protoplanetary disk. According to the analytic formula, the mass of formed protoplanets is proportional to $Q_{\rm RD}^{*}{}^{0.87}$: a factor of two difference in $Q_{\rm RD}^{*}$  directly affects the mass of protoplanets by a factor of 1.8. On the other hand, the final mass of protoplanets can be determined by a balance between the growth time of embryos and the mass depletion time in collision cascades using the simple fragmentation model. 
The mass depletion time is dominated by erosive collisions where the ejected mass is nicely scaled by $Q_{\rm R}/Q_{\rm RD}^{*}$.
Thus, the depletion time also depends on the value of $Q_{\rm RD}^{*}$ (In fact, the depletion time is proportional to $Q_{\rm RD}^{*}{}^{0.69}$ \citep{KT10}). 
As a result, the determination of the values of $Q_{\rm RD}^{*}$ can provide constraints on the formation of planetary bodies.

In this study, we showed the correspondence of $Q_{\rm RD}^{*}$ for the case of  purely hydrodynamic bodies.
To determine a more realistic value of $Q_{\rm RD}^{*}$ for various types of planetesimals, 
how $Q_{\rm RD}^{*}$ depends on material properties needs to be clarified. 
Therefore, we will investigate the effects of material properties in the future.

\acknowledgements
We thank Thomas Davison and an anonymous reviewer for their useful comments on our manuscript.
We gratefully acknowledge the developers of iSALE, including Gareth Collins, Kai W\"{u}nnemann, Boris Ivanov, Jay Melosh, Dirk Elbeshausen, and Thomas Davison.
This work was supported by JSPS KAKENHI Grant Numbers JP26287101 and JP15H03716.
H.K. was supported by Grant-in-Aid for Scientific Research (JP17K05632, JP17H01105, JP17H01103) and JSPS Core-to-Core Program ``International Network of Planetary Sciences''.
H.G. was also supported by JSPS KAKENHI Grant Number JP17H02990.

%%%%%%%%%%%%%%%%%%%%%%%%%%%%%%%%%%%%%%%%%%%%%%%%%%%%%%%%%%%%%%%%%%%%%%%%%%%%%%%%%%%%%%%%%%%
%%%%%%%%%%%%%%%%%%%%%%%%%%%%%%%%%%%%%%%%%%%%%%%%%%%%%%%%%%%%%%%%%%%%%%%%%%%%%%%%%%%%%%%%%%%%

\newpage

\clearpage

\clearpage

\clearpage
\newpage

\begin{threeparttable}[htb]
\begin{center}
\caption{iSALE input parameters} \label{table:input}
\begin{tabular}{lcccc}
%\tableline\tableline 
\hline
Description & \multicolumn{4}{c}{Values} \\
\hline
Cell per target radius ($n_{\rm tar}$) & 100 & 200 & 400 & 800 \\
Horizontal cells ($n_{\rm h}$) & 450 & 900 & 1800 & 3600 \\
Vertical cells ($n_{\rm v}$) & 450 & 900 & 1800 & 3600 \\
Setup type &  \multicolumn{4}{c}{PLANET\tnote{a}}  \\
Surface temperature [K] &  \multicolumn{4}{c}{293\tnote{b}}  \\
Gradient type  & \multicolumn{4}{c}{SELF\tnote{a}} \\
Gradient dimension  & \multicolumn{4}{c}{3} \\
Self-gravity accuracy parameter ($\theta$) & \multicolumn{4}{c}{1.0 or 0.5 ($R_{\rm tar}=300$ km)} \\
Self-gravity update frequency  & \multicolumn{4}{c}{10} \\
Volume fraction cutoff  & \multicolumn{4}{c}{$10^{-6}$\tnote{b}} \\
Density cutoff [kg/m$^{3}$] & \multicolumn{4}{c}{5\tnote{b}} \\
Velocity cutoff  & \multicolumn{4}{c}{1.697$\times v_{\rm imp}$\tnote{b}} \\
Linear term of artificial viscosity & \multicolumn{4}{c}{0.24\tnote{b}} \\
Quadratic term of artificial viscosity & \multicolumn{4}{c}{1.2\tnote{b}} \\
\hline
\end{tabular}
\begin{tablenotes}
\item[a] See \citet{C16} for detailed description.
\item[b] Default parameter values of the iSALE code (See \citet{C16}).
\end{tablenotes}
\end{center}
\end{threeparttable}

\begin{landscape}
\begin{table}[htb]
\begin{center}
\caption{Collision outcomes} \label{table:output}
\begin{tabular}{ccccp{1pt}ccp{1pt}ccp{1pt}cc}
\hline
\multicolumn{13}{c}{$R_{\rm tar}=100$km, $M_{\rm tar}=1.13\times10^{19}$kg,  $v_{\rm imp}=3$ km/s} \\
\hline
 \multirow{2}{*}{$M_{\rm imp}$[kg]}  &  \multirow{2}{*}{$Q_{\rm R}$[J/kg] }  &  \multicolumn{2}{c}{$n_{\rm tar}=100$} &&  \multicolumn{2}{c}{$n_{\rm tar}=200$} &&  \multicolumn{2}{c}{$n_{\rm tar}=400$} &&  \multicolumn{2}{c}{$n_{\rm tar}=800$}\\
\cline{3-4}   \cline{6-7} \cline{9-10} \cline{12-13}                  
                                                            &                                                              & $n_{\rm imp}$ & $M_{\rm ej}/M_{\rm tot}$ && $n_{\rm imp}$ & $M_{\rm ej}/M_{\rm tot}$ && $n_{\rm imp}$ & $M_{\rm ej}/M_{\rm tot}$ && $n_{\rm imp}$ & $M_{\rm ej}/M_{\rm tot}$ \\
\hline
 $3.1\times10^{16}$    & $1.23\times10^{4}$ &  14 & $1.61\times10^{-1}$ && 28 & $2.02\times10^{-1}$  && 56 & $2.21\times10^{-1}$  && 112 & $2.48\times10^{-1}$\\
 $3.82\times10^{16}$  & $1.51\times10^{4}$ &  15 & $2.25\times10^{-1}$  && 30 & $2.58\times10^{-1}$  && 60 & $3.32\times10^{-1}$  && 120 & $3.69\times10^{-1}$\\
 $4.63\times10^{16}$  & $1.83\times10^{4}$ &  16 & $2.91\times10^{-1}$  && 32 & $4.09\times10^{-1}$  && 64 & $4.6\times10^{-1}$    && 128 & $4.83\times10^{-1}$\\
 $5.56\times10^{16}$  & $2.19\times10^{4}$ &  17 & $4.18\times10^{-1}$  && 34 & $5.27\times10^{-1}$  && 68 & $5.67\times10^{-1}$  && 136 & $5.83\times10^{-1}$\\
 $6.6\times10^{16}$    & $2.59\times10^{4}$ &  18 & $5.6\times10^{-1}$   && 36 & $6.36\times10^{-1}$  && 72 & $6.69\times10^{-1}$  && 144 & $6.83\times10^{-1}$\\
 $7.76\times10^{16}$  & $3.04\times10^{4}$ &  19 & $6.7\times10^{-1}$   && 38 & $7.53\times10^{-1}$  && 76 & $7.74\times10^{-1}$  && 152 & $7.82\times10^{-1}$\\
 $9.05\times10^{16}$  & $3.54\times10^{4}$ &  20 & $8.07\times10^{-1}$  && 40 & $8.39\times10^{-1}$  && 80 & $8.63\times10^{-1}$  && 160 & $8.71\times10^{-1}$\\
 $1.05\times10^{17}$  & $4.09\times10^{4}$ &  21 & $8.88\times10^{-1}$  && 42 & $9.21\times10^{-1}$  && 84 & $9.33\times10^{-1}$  && 168 & $9.39\times10^{-1}$\\ 
\hline
\end{tabular}
\end{center}
\end{table}
\end{landscape}

\begin{threeparttable}
%\begin{table}[htb]
\begin{center}
\caption{Constants in erosive and disruptive collisions}\label{table:qg}
\begin{tabular}{lccc}
\hline
Constants & Water & Rock & Sand \\
\hline
$\mu$      & 0.55\tnote{a} & 0.55\tnote{b} & 0.41\tnote{c}  \\
$C_{0}$    & 1.5\tnote{a} & 1.5\tnote{b} & 0.55\tnote{c}  \\
$k$           & 0.2\tnote{a}  & 0.3\tnote{b} & 0.3\tnote{c}  \\
$\rho$ [kg/m$^{3}$]  & 1000\tnote{a}  & 3000\tnote{b} & 1600\tnote{c}  \\
$\Pi_{\rm G}^{*}/\psi$ & 15.4 & 10.3 & 26.9  \\
\hline
\end{tabular}
\begin{tablenotes}
\item[a] \citet{SH87,HH11}
\item[b] \citet{G63} 
\item[c] \citet{C99}
\end{tablenotes}
\end{center}
\end{threeparttable}

\begin{threeparttable}
\begin{center}
\caption{iSALE material parameters}\label{table:material}
\begin{tabular}{lc}
%\tableline\tableline 
\hline
Description & Values\tnote{a} \\
\hline
Poisson's ratio & 0.25 \\
Specific heat capacity [J/(kg K)]  & $10^{3}$ \\
Strength model & Rock\tnote{b} \\
Cohesion (undamaged) [Pa] &  $2.5\times10^{6}$ \\
Coefficient of internal friction (undamaged)  & $2.0$ \\
Limiting strength at high pressure (undamaged) [Pa] & $2\times10^{9}$ \\
Cohesion (damaged) [Pa]    & $10^{4}$ \\
Coefficient of internal friction (damaged)     & 0.4 \\
Limiting strength at high pressure (damaged) [Pa] & $2.5\times10^{9}$ \\
Damage model  & Ivanov\tnote{c} \\
Minimum failure strain  & $10^{-4}$ \\
Damage model constant & $10^{-11}$ \\
Threshold pressure for damage model [Pa] & $3\times10^{8}$ \\
\hline
\end{tabular}
\begin{tablenotes}
\item[a] Parameter values generated by iSALEMat.
\item[b] See \citet{I97} for detailed description.
\item[c] See \citet{C04} for detailed description.
\end{tablenotes}
\end{center}
\end{threeparttable}

\end{document}